\documentclass[10pt,a4paper]{article}
\usepackage[latin1]{inputenc}
\usepackage{amsmath}
\usepackage{amsfonts}
\usepackage{amssymb}
\usepackage{graphicx}
\usepackage{hyperref}
\title{The Causal Closure of Physics in Real World Contexts}

\author{George F R Ellis\\
Mathematics Department, University of Cape Town
}
\begin{document}
	\maketitle
	
\begin{abstract}
\noindent	The causal closure of physics is usually discussed in a context free way. Here I discuss it in the context of engineering systems and biology, where strong emergence takes place due to a combination of upwards emergence and downwards causation \cite{Ellis_2020_response}. Firstly, I show that causal closure is strictly limited in terms of spatial interactions because these are  cases that are of necessity strongly interacting with the environment. Effective Spatial Closure holds \textit{ceteris parabus}, and can be violated by Black Swan Events. Secondly, I show that causal closure in the hierarchy of emergence is a strictly interlevel affair, and in the cases of engineering and biology encompasses all levels from the social level to the particle physics level. However Effective Causal Closure can usefully be defined for a restricted set of levels, and one can experimentally determine Effective Theories that hold at each level. This does not however imply those effective theories are causally complete by themselves. In particular, the particle physics level is not causally complete by itself in the contexts of solid state physics (because of interlevel wave-particle duality), digital computers (where algorithms determine outcomes), or biology (because of time dependent constraints). Furthermore Inextricably Intertwined Levels occur in all these contexts.
\end{abstract} 
\tableofcontents
\newpage
\section{The Context}\label{sec:Intro}
It is often supposed that causal closure occurs at the micro level in physical systems, and hence prevents the occurrence of strong emergence 
  because the macrostate supervenes on the microstate \cite{Kim 1998} \cite{Kim 1999}.  This is discussed in   
  \cite{Clayton and Davies 2006} \cite{Hohwy and Kallestrup}  \cite{Macdonald and Macdonald 2010} 
  and \cite{Handbook_of_Emergence}. 
   In contrast,  \cite{Butterfield 2011} shows by careful philosophical argument that 
 \begin{quote}
 	``\textit{One can have emergence with
 	reduction, as well as without it; and emergence without supervenience, as well as
 	with it.}''
 \end{quote}
Here I want to examine the issue in a different way, by dealing in some detail with the hierarchical nature of emergence in real world contexts: the cases of engineering, based in the underlying solid state physics,  and biology, based in the underlying  molecular biology, in turn based in the underlying physics. 
The context is my paper \cite{Ellis_2020_response} that establishes that strong emergence does indeed take place in both those cases, so that the argument against strong emergence has to be wrong in those contexts.  

 \cite{Ellis_2020_response} examines the issue of strong emergence of properties \textbf{P} of macrodynamics \textbf{M} out of the underlying microdynamics \textbf{m} in the context of condensed matter physics and biology. Following Anderson's lead \cite{Anderson_72} \cite{Anderson_94}  
that symmetry breaking is at the heart of emergence, its method was to  identify five different kinds of symmetry breaking occurring in different contexts (Section 2.3 of \cite{Ellis_2020_response}) and then to trace how broken symmetry states at the macro and micro levels interact with each other. 

\begin{itemize}
	\item The microscale dynamics \textbf{m}, based in the Laws of Physics $L$, obeys symmetries \textbf{S}: \textbf{S}(\textbf{m})=\textbf{m}. \href{https://en.wikipedia.org/wiki/Spontaneous_symmetry_breaking}{Spontaneous Symmetry Breaking} \textbf{SSB}(\textbf{M}) leads to symmetry breaking of the macro scale dynamics \textbf{M} through the emergence process \textbf{E} whereby \textbf{M} emerges from the microscale dynamics \textbf{m}. This is weak emergence $\textbf{E}: \textbf{m} \rightarrow \textbf{M}:  
	 \textbf{S}(\textbf{M}) \neq \textbf{M}$.  
\item This spontaneously broken macro state \textbf{M} reaches down to 
create \href{https://en.wikipedia.org/wiki/Quasiparticle}{quasiparticles} such as \href{https://en.wikipedia.org/wiki/Phonon}{phonons} at the micro level, which play a key dynamical role at that level.\footnote{See the discussion in Section \ref{sec:Intertwined_computers} below.}  The base microdynamics \textbf{m} is altered to produce an effective microdynamics \textbf{m'} which breaks the symmetry of the underlying physical laws $L$. Thus  $\textbf{m} \rightarrow \textbf{m'}: \textbf{S}(\textbf{m'})\neq \textbf{m}$.   
	\item To derive correctly the properties of macro dynamics \textbf{M} from the micro dynamical level, you must  coarse grain the effective theory \textbf{m'} rather than \textbf{m}.
	\item Thus strong emergence takes place in this case: you cannot even in principle derive the macrodynamics \textbf{M} from the microdynamics \textbf{m} in a strictly bottom up way, because \textbf{m} satisfies the symmetry \textbf{S} and \textbf{M} does not.\footnote{Unless the coarse-graining operation \textbf{C} breaks the symmetry \textbf{S}: then $\textbf{CS} \neq \textbf{SC}$.}  
\end{itemize}
As a consequence (\cite{Ellis_2020_response}: Section \textbf{4.4}), 
in the case of solid state physics, the underlying microphysics \textbf{m} cannot be causally complete, because by itself it cannot lead to the emergence of known properties of solids such as electrical conductivity. The lower level physics only gives the correct outcome when modified by inclusion of terms \textbf{a(M)} arising from the higher level state \textbf{M} (so $\textbf{S(a)}\neq \textbf{a}$). 
The same is true for living systems.
That is, in both these cases, causal completeness is  only attained by considering both the low level properties \textbf{m} and the higher level properties \textbf{M} (which lead to the alteration $\textbf{m} \rightarrow \textbf{m'}$) together. The real causally closed system comprises both those levels.\\

The aim of this paper is to extend that result by investigating causal closure of physics in terms of determining dynamic properties\footnote{I am using the classification in Section 2.2 of \cite{Ellis_2020_response}.}  \textbf{P(d)} of entities in engineering and biological contexts (I use the word closure rather than completeness for reasons that become apparent below as the theme develops). A separate very interesting project would extend this to considering the dynamic emergence of entities \textbf{E(d)} over time, see e.g.  
\cite{Carroll 2005}.

  Section \ref{sec:hierarchy} sets the context for the discussion, which is the hierarchies of emergence in the cases of engineering and the life sciences  respectively (\textbf{Table 1}). It introduce the idea of an \textit{Effective Theory} $\textbf{ET}_\textbf{L}$ at each level \textbf{L}, and discusses bottom up and top down causation in the hierarchy of emergence (Section  \ref{sec:upward_downward}. 
  
 Section \ref{sec:Domains_of_interest} introduces the idea of a  \textit{Domain of Interest} \textbf{(DOI)}, and the concept of \textit{Effective Spatial Closure} (Section  \ref{sec:open_controlled}). It is shown that in terms of spatial interactions, causal closure in engineering and biology only holds \textit{ceteris parabus}. Yes of course philosophers know that this is the case; the point is that it has real consequences in real world contexts.
 
 Section  \ref{sec:Levels of Interest}  defines  \textit{Levels of Interest} \textbf{(LOI)} and Section \ref{sec:LOI_restricted} introduces the need for \textit{Restricted Domains of Interest}. 
 Section \ref{sec:interlevel closure} introduces \textit{Interlevel Causal Closure} in the case of biology (Section  \ref{sec:causal_closure_biology}), 
 Section  \ref{sec:causal_closure_computers} extends this to the case of digital computers and physics. 
 
Section \ref{sec;conclude} summarizes the main results of this paper, emphasizes that unavoidable unpredictability also undermines causal closure, and comments on ways people ignore the issues raised in this paper.  

The novel concepts introduced  are \textit{Effective Theories} $\textbf{ET}_\textbf{L}$ (Section   \ref{sec:equal_valid}), \textit{Effective Spatial Closure} (Section \ref{sec:open_controlled}), \textit{Levels of Interest} (\textbf{LOI}) (Section \ref{sec:Levels of Interest}),  \textit{Effective Causal Closure}  (\textbf{ECC}) (Section \ref{sec:causal_closure_biology}), and 
\textit{Inextricably Intertwined Levels} (\textbf{IIL}) (Section  \ref{sec:Intertwined_Biology}).
\section{The Hierarchy of Emergence}\label{sec:hierarchy}
The context of the discussion is the hierarchy of emergence. As stated by \cite{Anderson_72},
\begin{quote}
	\textit{``At each level of complexity, entirely new properties appear ... At each stage entirely new laws, concepts, and generalizations are necessary.''}
\end{quote}
In this section, I present the nature of the hierarchy (Section \ref{sec:hierarchy_nature}), and comment on Effective Theories and the Equal Validity of Levels (Section \ref{sec:equal_valid}. The latter is enabled by a combination of upward and downward causation (Section \ref{sec:upward_downward}), with the key feature of multiple realisability of higher levels in terms of lower level states (Section   \ref{sec_multiple_realise}). The crucial relation between Effective Theories and Causal Closure is briefly  commented on in Section \ref{sec:ET_CC}. Going into more detail as regards the hierarchical structure, it is modular (Section \ref{sec:modular}), with the modules forming networks (Section \ref{sec:networks}).
\subsection{The nature of the hierarchy}\label{sec:hierarchy_nature}
The emergent hierarchy  is shown in \textbf{Table 1} for the cases of engineering  on the left and life sciences, in  particular the case of humanity, on the right. The left hand side represent \textit{The Sciences of the Artificial} as discussed by \cite{Simon 2019}. The right hand side represents the structures and processes of biology, as discussed by \cite{Campbell and Reece 2005}.

\vspace{0.1in}
	\begin{tabular}{|c|c|c|c|}
	\hline \hline
	& \textbf{Engineering} & \textbf{Life Sciences} \\	
	\hline Level 9 (\textbf{L9}) & Environment &  Environment \\ 
	\hline 
	Level 8	(\textbf{L8}) & Sociology/Economics/Politics &  Sociology/Economics/Politics \\ 
	\hline 
	Level 7	(\textbf{L7}) & Machines &  Individuals \\ 
	\hline 
	Level 6	(\textbf{L6}) & Components &  Organs \\ 
	\hline 
	Level 5	(\textbf{L5}) & Devices & Cells  \\ 
	\hline 
	Level 4 (\textbf{L4})	& Crystals & Biomolecules \\ 
	\hline 
	Level 3 (\textbf{L3}) & Atomic Physics	& Atomic Physics
	\\ 
	\hline 
	Level 2	(\textbf{L2}) & Nuclear Physics & Nuclear Physics  \\ 
	\hline 
	Level 1	(\textbf{L1}) &  Particle Physics & Particle Physics  \\ 
	\hline \hline
\end{tabular} 
\vspace{0.1in}

\noindent \textbf{Table 1}: \textit{The emergent hierarchy of structure and causation for engineering  (left) and life sciences (right)} (developed from \cite{Ellis_2016} and \cite{Ellis_Drossel_computers}).\\

\noindent This hierarchy has important aspects.
\begin{itemize}
	\item The bottom three levels \textbf{L1}-\textbf{L3} are the same on both sides. This is one of the great discoveries science has made: inanimate matter and living matter are made of the same stuff at the bottom. Electrons are 
	 at level \textbf{L3}, interacting with the nucleus.
	\item The atomic level \textbf{L3} is where new properties emerge out of the underlying physics, as characterised by the \href{https://en.wikipedia.org/wiki/Periodic_Table_of_the_Elements}{Periodic Table of the Elements}, another great discovery.
	\item The components enabling complexity to arise occur at level \textbf{L4}. Both  \href{https://en.wikipedia.org/wiki/Semiconductor}{semiconductors} and  \href{https://en.wikipedia.org/wiki/Metals}{metals} are crystals, and they are the key components of machines.\footnote{\href{https://en.wikipedia.org/wiki/Amorphous_solid}{Amorphous materials} such as glasses may also occur, but they do not play a key role in the dynamic emergence of properties \textbf{P}(\textbf{d}) in machines.} Solid State Physics covers levels \textbf{L3}-\textbf{L4}. Biomolecules such as \href{https://en.wikipedia.org/wiki/DNA}{DNA}, \href{https://en.wikipedia.org/wiki/RNA}{RNA}, and \href{https://en.wikipedia.org/wiki/Protein}{proteins}  are the foundations of biological emergence. 
	\item Level \textbf{L5} is where the basic units of complexity arise, showing functional emergent properties, and being the basis for building complex entities. On the machines side, these are devices such as \href{https://en.wikipedia.org/wiki/Transistor}{transistors}, \href{https://en.wikipedia.org/wiki/Light_emitting_diodes}{light emitting diodes}, \href{https://en.wikipedia.org/wiki/Photodetector}{photodetectors}, and \href{https://en.wikipedia.org/wiki/Lasers}{lasers}. On the life sciences side, they are \href{https://en.wikipedia.org/wiki/Cell_(biology)}{cells}: the basic building block of life, which come in many different types. This is the lowest level  where the processes of life occur, entailing metabolism and information processing  \cite{Hartwell_et_al_1999}.
	\item Level \textbf{L6} is where on the machine side, devices are integrated into functional units, such as the \href{https://en.wikipedia.org/wiki/CPU}{Central Processing Unit} in a \href{https://en.wikipedia.org/wiki/Microprocessor}{Microprocessor}\,, which is itself a device. On the life sciences side,  \href{https://en.wikipedia.org/wiki/Organ_(anatomy)}{organs} comprising \href{https://en.wikipedia.org/wiki/Physiology}{physiological systems} occur.  
	\item Level \textbf{L7} is where functional units occur that have an integrity of their own: they are effectively causally closed systems imbedded in a larger environment. On the engineering side, they are \href{https://en.wikipedia.org/wiki/Machine}{machines} built to carry out some purpose, such as aircraft or  digital computers or particle colliders. On the life sciences side, they are \href{https://en.wikipedia.org/wiki/Individual}{individuals} with autonomy of action.  
	\item Level \textbf{L8} is the same on both sides. Both machines and individual human beings exist in the context of a \href{https://en.wikipedia.org/wiki/Society}{society} with social, economic, and political aspects,  which sets the stage for their existence and functioning. 
	\item Finally, level \textbf{L9} is again the same on both sides. It reflects the fact that each society exists in a  \href{https://en.wikipedia.org/wiki/Natural_environment}{natural environment} with both ecological and geophysical aspects.  
\end{itemize}
Note that this Table has chosen a particular set of levels to represent causation all the way from the Particle Physics Level \textbf{L1} to the Environmental Level \textbf{L9}. However most scientific studies will be interested in a much more restricted sets of levels: the \textit{Levels of Interest} (\textbf{LOI}s) discussed in Section \ref{sec:Levels of Interest}. Given such a choice, one will in general use a more fine-grained set of levels than represented in \textbf{Table 1}. Thus for example if the \textbf{LOI} is (\textbf{L4}\textbf{-}\textbf{L6}), one might divide that range into a finer set of sublevels.
 
\subsection{Effective Theories and the Equal Validity of Levels}\label{sec:equal_valid}
It is a common belief that the lower levels are more real than the higher levels, because bottom up causation from the lower to higher levels is the source of higher level properties.  Arthur Eddington 
 in 
  \textit{On the Nature of the Physical World} (\cite{Eddington_29}:5-12) muses on the dual (solid macroscopic/atomic microscopic) nature of his writing desk, and concludes (page 10) that because of the  scientific world view,
\begin{quote}
	\textit{``The external world of physics has thus become a world of shadows. In removing our illusions we have removed the substance, for indeed we have seen that substance is one of our great illusions.''}
\end{quote}
However this view is subject to dispute.  Richard Feynman in his book \textit{The Character of Physical Law} (\cite{Feynman 2017}:125-126) considers whether one level or another is more fundamental, and using a religious metaphor, argues that `\textit{the fundamental laws are no nearer to God than emergent laws}'\,.\footnote{This passage is quoted in fulll in  \cite{Ellis_2016}, pages 454-455.}  Phil Anderson arguably had a similar view. Sylvan Schweber commented as follows \cite{Schweber 1993}:
\begin{quote}
\textit{``Anderson  believes in  emergent laws.   He holds the view that each level has its  own ``fundamental'' laws and its own ontology. Translated into the language of particle physicists, Anderson would say each level has its effective Lagrangian and its set of quasistable particles. In  each level the effective Lagrangian - the ``fundamental'' description at that level -  is  the best  we can  do.}  
\end{quote}
Thus this does not recognize any level as more fundamental than any other.

Recently, Denis Noble has proposed a ``\textit{Principle of Biological Relativity}'' \cite{Noble 2012}: all levels one deals with in studying emergence in biology are equally valid, there is no privileged level of causation. 

\paragraph{Effective Theories} 
A good way to express this is that there is a valid \textit{Effective Theory}\footnote{Not to be confused with an \textit{Effective Field Theory} (EFT), see \cite{Castellani_02},  \cite{Burgess_2007}, 
	\cite{Hartmann 2001}, 
	 which is a special case of an ET. Note that  EFTs such as in \cite{LM} cannot deal with emergence in solid state physics, as they do not allow for symmetry breaking.} (ET) at each level. Elena Castellani  gives this definition \cite{Castellani_02}: 
\begin{quote}
	``\textit{An effective theory (ET) is a theory which `effectively' captures what is physically relevant in a given domain, where `theory' is  a set of fundamental equations (or simply some Lagrangian) for describing some entities, their behaviour and interactions... More precisely,  an ET is an appropriate description of the important (relevant) physics in a given region of the parameter space of the physical world.}
\end{quote}
In parallel to the way the functioning of the Laws of Phyiscs was sketched in \cite{Ellis_2020_response}, one can characterise an Effective Theory $\textbf{ET}_\textbf{L}$ valid at some level \textbf{L} as follows; 
\begin{quote}
\textit{An \textbf{Effective Theory} $\textbf{ET}_\textbf{L}$ at a level \textbf{L} is a reliable relation between initial conditions described by effective variables $v_\textbf{L} \in \textbf{L}$ and outcomes $o_\textbf{L}\in \textbf{L}$:}
\begin{equation}\label{eq:effective_laws}
\textbf{ET}_\textbf{L}: v_\textbf{L} \in \textbf{L} \rightarrow \textbf{ET}_\textbf{L}[v_\textbf{L}] = o_\textbf{L} \in \textbf{L}
\end{equation}\label{eq:ET}
\textit{in a reliable way, whether $\textbf{ET}_\textbf{L}$ is an exact or statistical law.}
\end{quote}
It is important to note that an effective theory may have a randomisation element ${\cal R}$:
\begin{equation}\label{eq:random effects}
 {\cal R}: v_\textbf{L} \in \textbf{L} \rightarrow {\cal R}(v_\textbf{L}) = v'_\textbf{L} \in \textbf{L}
\end{equation}
where  ${\cal R}$ might for example produce a Gaussian distribution. 

\paragraph{Equal Causal Validity} In terms of Effective Theories for emergent properties \textbf{P(d)},  Noble's principle \cite{Noble 2012}  as extended in \cite{Ellis_2020_response} can be restated:
\begin{quote}
	\textbf{Equal Causal Validity}: \textit{\textbf{Each emergent level \textbf{L} in the hierarchy (characterised as in Table 1) represents an Effective Theory $\textbf{ET}_\textbf{L}$, so each level is equally valid in a causal sense}.}
\end{quote}
This implies no level is a fundamental level with priority over the others, and particularly there is not a primary one at the bottom level. This is just as well, because  there is no well-established bottom-most physical level to which physics can be reduced \cite{Murugan Weltmann Ellis 2012}.  Every emergent level equally 
 represents an effective theory.\footnote{While  Luu and  Mei{\ss}ner are critical of my claims on emergence \cite{LM}, they agree on this point.}

\subsection{Upward and Downward Causation}\label{sec:upward_downward}
Equality of validity of effective theories at every level is possible because causation is not just bottom-up. Rather higher level properties \textbf{P(d)} are linked to lower levels by a combination of upwards and downwards causation (\cite{Noble 2012}, \cite{Ellis_2016}, \cite{Ellis_2020_response}), which enables emergence of effective laws at each level. 

\paragraph{Upwards emergence} This has two different aspects (\cite{Ellis_2020_response}:\S 1.1). 

First there is the Emergence \textbf{E} of a macro system  from its components. In terms of levels, this corresponds to creation of a higher level \textbf{LN} from a lower level \textbf{Ln}: that is,  $\textbf{Ln} \rightarrow \textbf{LN},\textbf{ N} > \textbf{n}$. This may lead to topological non-trivial states emerging such as networks, or Quantum Entanglement may take place .
The issue of  \href{https://en.wikipedia.org/wiki/Phase_transition}{phase transitions} is important here. First  order phase transitions  occur when Spontaneous Symmetry Breaking \textbf{SSB} occurs leading to the emergent level \textbf{ET} having lower symmetries than the underlying \textbf{ET}. In terms of the associated micro dynamics \textbf{m} and macrodynamics \textbf{M}, if \textbf{S} is the symmetry set of \textbf{m}, then
\begin{equation}\label{eq:upwards emergence}
\{\textbf{E}: \textbf{m} \rightarrow \textbf{M},\,\, \textbf{S(m) = \textbf{M}}\}\,\, \Rightarrow \,\,\textbf{S(\textbf{M})} \neq \textbf{M}.
\end{equation}

Second there is emergence \textbf{P} of properties of the emergent level \textbf{LN} out of properties of the underlying constituent level  \textbf{Ln} once \textbf{LN} has come into existence. This corresponds to emergence of a higher level $\textbf{ET}_\textbf{L}$ out of a lower level one. Some form of coarse graining \textbf{C} of properties may suffice if the higher and lower levels have the same symmetries \textbf{S}, but not if their symmetries are different due to \textbf{SSB} (see \cite{Ellis_2020_response}).

\paragraph{Downward causation} A classification of different types of downward causation was given in \cite{Ellis 2012}, \cite{Ellis_2016}. Here I will rather approach the issue  from an $\textbf{ET}_\textbf{L}$ viewpoint. There are essentially two kinds of downwards effects that can happen: downward alteration of lower level dynamics \textbf{L} via either constraints or effective potentials, and downward alteration of  dynamics at level \textbf{L} by altering the set of lower level  variables.

\paragraph{Constraints and Effective Potentials} The way downward causation by constraints works  is that the outcomes \textbf{P(d)} at Level \textbf{L} depend on constraints ${\cal C}_\textbf{LI}$ at the level \textbf{L} arising from conditions at a Level of Influence $\textbf{LI}$. Thus when interlevel interactions are taken into account, relation (\ref{eq:effective_laws}) is modified (see (30) in \cite{Ellis_2020_response}) to 
\begin{equation}\label{eq:constraints}
\textbf{ET}_\textbf{L}({\cal C}_\textbf{LI}) : v_\textbf{L} \in \textbf{L} \rightarrow \textbf{ET}_\textbf{L}({\cal C}_\textbf{LI}) [v_\textbf{L}] = o_\textbf{L} \in \textbf{L}
\end{equation}
Essentially the same holds if the effect of the Level of Influence \textbf{LI} on the level \textbf{L} is expressed in terms of an effective  potential $V(v_\textbf{LI})$ at level \textbf{L} (see (9) in \cite{Ellis_2020_response}). Thus constraints act as causes \cite{Juarrero 2002}.

The constraints ${\cal C}_\textbf{LI}$ may be time independent: 
$\partial \,{\cal C}_\textbf{LI}/\partial t = 0
$ in which case they are structural constraints; or they may be time dependent: ${\cal C}_\textbf{LI} = {\cal C}_\textbf{LI}(t), \partial {\cal C}_\textbf{LI}/\partial t \neq 0$, in which case they are signalling or controlling constraints. An important case is feedback control (engineering), which is essentially the same as homeostasis (biology). Then the constraints  ${\cal C}_\textbf{LI}(t)$ depend on goals $G_ \textbf{LI}$ valid at level \textbf{L} but set at the Level of Influence \textbf{LI}. Similarly the potential $V(v_\textbf{LI})$ may depend on time-dependent variables $v_{\textbf{LI}}(t)$ at the Level of Influence \textbf{CL}. Then $\partial v_\textbf{LI}(t)/\partial t \neq 0 \Rightarrow \partial V(v_\textbf{LI})/\partial t \neq 0$. 
In both cases the level \textbf{L} is no longer causally complete on its own; at a minimum, only the combination \{\textbf{L},\textbf{LI}\} of levels can be causally complete.

\paragraph{Altered variables}
 The causal effect due to the level of influence \textbf{LI} may rather be due to changes in the variables $v_\textbf{L}$ at level \textbf{L} due to variables $v_{\textbf{LI}}$ at the higher level \textbf{LI}:
 \begin{equation}\label{eq:altered-variables}
 \textbf{ET}_\textbf{L}({ v}_\textbf{LI}) : \{v_\textbf{L}\} \in \textbf{L} \rightarrow \textbf{ET}_\textbf{L}({ v}_\textbf{LI})[v_\textbf{L}] =  \{v'_{\,\,\,\textbf{L}}\} \in \textbf{L}
 \end{equation}
 where the new set $\{v'_{\,\,\,\textbf{L}}\}$ of effective variables at level \textbf{L} may be smaller, larger, or altered.

They are \textit{smaller} if they are changed by deleting lower level elements. This occurs when \textit{Downward Causation by Adaptive  Selection} takes place, altering or deleting selected lower level elements according to some selection criterion \textbf{c}. This enables alteration of structures and functions  at level \textbf{L} so as to meet new challenges at level \textbf{LI}. This plays an important role in enabling by organisms to have agency and choice, enabled by stochasticity, as explained in \cite{Noble and Noble 2018 Stochasticity}:
\begin{quote}
	"\textit{Choice in the behavior of organisms involves novelty, which may be unpredictable. Yet in retrospect,
		we can usually provide a rationale for the choice. A deterministic view of life cannot explain
		this. The solution to this paradox is that organisms can harness stochasticity through which they
		can generate many possible solutions to environmental challenges. They must then employ a comparator
		to find the solution that fits the challenge. What therefore is unpredictable in prospect can
		become comprehensible in retrospect. Harnessing stochastic and/or chaotic processes is essential to
		the ability of organisms to have agency and to make choices''}
\end{quote}

They are \textit{larger} if for example one has downward creation of quasiparticles such as phonons via interlevel wave-particle duality (\cite{Ellis_2020_response} and Section  \ref{sec:Intertwined_computers}), which  underlies the properties of metals and semi-conductors.  This is what  Carl Gillett   
calls a ``Foundational Determinative Relation'' \cite{Gillett}). 

The are \textit{altered} if the number is the same but the properties of an element changes. When they are bound in an emergent complex their own properties may change (for example, neutrons decay in 11 minutes when free but last for billions of years when bound in a nucleus), or their interactions with external entities may change (for example electrons bound in an atom interact with light quite differently than a free electron does).

Downward causation is related to Aristotle's \textit{Formal Cause}, see \cite{Tabaczek 2013}, but I will not follow that strand here. To give these rather abstract statements flesh, 
 see many examples given  in \cite{Noble2008_Music}, 
   \cite{Ellis_2016}. Downward causation in relation to the  key physics-chemistry link is discussed in \cite{Luisi}.
\subsection{Multiple Realisability}\label{sec_multiple_realise}
 Multiple realisability of higher level variables at lower levels plays a key role in downward causation \cite{Menzies 2003}. Any particular higher level state can be realised in a multiplicity of ways in terms of lower level states. In an engineering or biological cases, a high level need determines the high level effective function that needs to be realised and thus the high level structure that fulfills it. This higher structure and function is then realised by suitable lower level structures and functions,  but there are billions of ways this can happen

It does not matter which of the equivalence class of lower level realisations is used to fulfill the higher level need,  as long as it is indeed fulfilled. Consequently you cannot even express the dynamics driving what is happening  in a sensible way at a lower level. 

The issue is not just the huge number lower level entities involved in realising a higher level systems, as characterised by \href{https://en.wikipedia.org/wiki/Avogadro_constant}{Avagadro's Number} 
It is the huge different numbers of ways combinations of lower level entities can represent a single higher level variable. Any one of the entire equivalence class at the lower level will do. Thus it is not the individual variables at the lower level that are the key to what is going on: it is the equivalence class to which they belong. But that whole equivalence class can be describer by a single variable at the macro level, so that is the real effective variable in the dynamics that is going on. This is a kind of interlevel duality:
\begin{equation}\label{key}
\{v_\textbf{L} \in \textbf{L} \}\Leftrightarrow \{v_\textbf{i}: v_\textbf{i} \in E_{\textbf{L-1}}(v_\textbf{L-1})\in \textbf{(L-1)}\} 
\end{equation}
where $E_{\textbf{L-1}}(v_\textbf{L-1})$ is the equivalence class of variables $v_\textbf{L-1}$ at Level $\textbf{L-1}$ corresponding to the one variable $v_\textbf{L}$ at Level \textbf{L.} The effective law $\textbf{EF}_\textbf{L}$ at Level \textbf{L} for the (possibly vectorial or matrix) variables $v_\textbf{L}$ at that level is equivalent to a law for an entire equivalence class $E_{\textbf{L-1}}(v_\textbf{L-1})$ of variables at Level \textbf{L-1}. It does not translate into an Effective Law for natural variables $v_{\textbf{L-1}}$ \textit{per se} at Level \textbf{L-1}.

\subsection{Effective Theories and Causal Closure}\label{sec:ET_CC} 
It is important to note the following: one establishes the validity of an $\textbf{ET}_\textbf{L}$ for some chosen level \textbf{L} by doing experiments or making observations on phenomena occurring at that level. This involves the experimenter intervening at the level \textbf{L}, hence it is an interlevel interaction. For example a particle physics experiment considers Effective Laws at level \textbf{L1} but involves scientists at level \textbf{L7} and organisations at level \textbf{L8} acting down to affect things at level \textbf{L1}. 
Consequently, one can make the following important observation:
\begin{quote}
	\textbf{Existence and functioning of Effective Theories $\textbf{ET}_\textbf{L}$ at level     \textbf{L} does not necessarily imply causal closure of Level \textbf{L}. }
\end{quote} 
The issue is what determines  constraints ${\cal C}$, potentials $V$, and effective variables $v'_{\textbf{LI}}$ that may occur at that level. They may be influenced by other levels. That is what Sections \ref{sec:interlevel closure} and \ref{sec:causal_closure_computers} are about. Determining an effective law at level \textbf{L} involves other levels then \textbf{L}. 

\subsection{Modular structure}\label{sec:modular}
Looking in more detail at the hierarchy, it is 
a hierarchy made of modules (this section) which form networks (next section). It is a modular hierarchy for very good reasons.

\paragraph{Five principles of complex structure } (\cite{Booch 2006}:\S1.3) gives five principles of complex structure, developing from \cite{Simon 2019}, starting from the idea 
\begin{quote}
	``\textbf{The Role of Decomposition}: 
	The technique of mastering complexity has been known since ancient times:
	\textit{divide et impera} (divide and rule)''
\end{quote}
The five principles, applicable to both engineering and biology, are stated by him to be,
\begin{enumerate}
	\item \textbf{Hierarchic Structure}:  \textit{Frequently, complexity takes the form of a hierarchy, whereby a complex system
	is composed of interrelated subsystems that have in turn their own subsystems,
	and so on, until some lowest level of elementary components is reached}
	\item \textbf{Relative Primitives}: \textit{The choice of what components in a system are primitive is relatively arbitrary
	and is largely up to the discretion of the observer of the system.}
	\item \textbf{Separation of Concerns}: \textit{Intracomponent linkages are generally stronger than intercomponent linkages.
	This fact has the effect of separating the high-frequency dynamics of the components,
	involving the internal structure of the components, from the low frequency
	dynamics, involving interaction among components}.
	\item \textbf{Common Patterns} \textit{Hierarchic systems are usually composed of only a few different kinds of subsystems
	in various combinations and arrangements.}
	\item \textbf{Stable Intermediate Forms} \textit{A complex system that works is invariably found to have evolved from a simple
	system that worked. . . . A complex system designed from scratch never works
	and cannot be patched up to make it work. You have to start over, beginning with
	a working simple system.}
\end{enumerate}
This underlies existence of levels such that each level is equally causally effective (\S \ref{sec:equal_valid}). Booch says ``\textit{Different
objects collaborate with one another through patterns of interaction that we call mechanisms}''. These are what I am calling Effective Theories (ETs). The objects that collaborate are modules.

\paragraph{Modules} Modularity is the property of a system that has been decomposed into a set of
cohesive and loosely coupled modules (\cite{Booch 2006}:56). They can be represented by \textit{Abstractions}, where
``\textit{An abstraction denotes the essential characteristics of an object that distinguish it
	from all other kinds of objects and thus provide crisply defined conceptual
	boundaries, relative to the perspective of the viewer.
	An abstraction focuses on the outside view of an object and so serves to separate
	an object's essential behavior from its implementation.}''
(\cite{Booch 2006}:44-50). They involve  \textit{Encapsulation} (\cite{Booch 2006}:50-53), that is, the internal details of the module's workings are hidden from the external world), and \textit{Multiple realisability}: the required module functioning can be fulfilled in many ways by its internal structure and variables. Hierarchy is a ranking or ordering of abstractions (\cite{Booch 2006}:58).

\subsection{Networks}\label{sec:networks}
A feature of particular interest is that emergent systems may give rise to Effective Theories that involve topological constraints. Indeed this happens quite often because emergent complexity in both engineering and biology often involves interaction networks, and a key feature of such networks is their topological connectivity, described by \href{https://en.wikipedia.org/wiki/Graph_theory}{graph theory}. Thus for example Arthur Peacocke points out that \begin{quote}
	``\textit{In electrical circuit theory there are certain topological constraints, the boundary conditions that one element imposes on another} (\cite{Peacocke 1990}:74)
\end{quote}   
They obviously have strongly emergent properties: their functioning  does not follow from any local characteristics of the elements that make up the circuit. The electric light won't work until you change its open circuit topology (isomorphic to an open interval) when the switch is off, to a closed topology (isomorphic to a circle) when the switch is on.   This macro event than reaches done to alter the flow of billions of electrons at the micro level.

Networks can be physical networks, or interaction networks. 

\paragraph{Physical Networks} 
\textit{Physical networks} are embodied in physical links between nodes,  which constrain what interactions can take place by dictating what nodes can interact with what other nodes. Thus physical networks in fact create interaction networks by constraining interactions between  links.  This is the key structure-function relationship of engineering and biology.
Examples 
in engineering are 
\href{https://en.wikipedia.org/wiki/Computer_architecture}{computer architecture} \cite{Tannenbaum},
  \href{https://en.wikipedia.org/wiki/Computer_network}{computer networks} \cite{Kurose 2005}, and \href{https://en.wikipedia.org/wiki/Artifiial_neural_network}{artificial neural networks} \cite{Jain 1996}. 
The case of importance in biology is 
the nervous system \cite{Guyton 1977} and \href{https://en.wikipedia.org/wiki/Neural_network}{neural networks} \cite{Haykin 1994}.

\paragraph{Interaction Networks} 
\textit{Interaction networks} occur due to the presence of a variety of reagents that selectively interact with each other. This requires firstly a container that keeps the reagents within interaction distance of each other, rather than just diffusing away, and second the presence of an appropriate set of reagent that do indeed interact with each other. A key role is then played by selectively letting specific reagents enter or exit the container so as to control their interaction densities. 

\textit{On a large scale}, examples of importance in engineering are
 purification plants,  chemical engineering reaction vessels, water treatment plants,
 sewage treatment plants .
In biology, they arguably  are the  \href{https://en.wikipedia.org/wiki/Endocrine_system}{endocryne system}, controlling signalling, and the \href{https://en.wikipedia.org/wiki/Human_digestive_system}{digestive system}, controlling metabolism at the systems level \cite{Guyton 1977}, and on a larger scale, ecological networks \cite{Junker and Schreiber 2011}.   

\textit{On a small scale}, there are many interaction networks in cell biology \cite{Buchanan 2010}. These are crucially dependent  on the existence of cells bounded by cell walls, that serve as the necessary reaction containers. They have ion channels imbedded in those walls that control movement of ions in and out of the cells, and molecular channels controlling movement of molecules in and out. They include
\begin{itemize}
	\item Gene regulatory networks \cite{Carroll 2005} \cite{Wagner 2014} \cite{Junker and Schreiber 2011}, also known as transcription networks \cite{Alon 2019}.
	\item Metabolic networks \cite{Buchanan 2010} \cite{Wagner 2014}\cite{Junker and Schreiber 2011}, 
	\item Cell signalling networks \cite{Berridge} \cite{Buchanan 2010} 
	\item Protein interaction networks\cite{Buchanan 2010} \cite{Junker and Schreiber 2011}
	\item Signal transduction networks \cite{Buchanan 2010} \cite{Junker and Schreiber 2011} 
\end{itemize}
These networks are the heart of cell biology \cite{Loewenstein 1999} and underlie how information flows and logic underlie biological functioning as emphasized by  \cite{Nurse 2008} \cite{Ellis_Kopel_2019} \cite{Davies 2019}. 

\paragraph{Networks and Hierarchy}
Networks may have a hierarchical character  in that subnetworks can often be identified within an overall network, and so define levels within the network \cite{Ravasz et al 2002} \cite{Papin et al 2004} \cite{Wuchty et al 2006}.
This is an interesting topic I will not develop further here 
except to remark that firstly, subnetworks include network motifs \cite{Mile Alon 2002} \cite{Alon 2019}, which are small subnetworks of particular functional significance. For example they include the \textit{autoregulation motif}, which is nothing other than feedback control (\cite{Alon 2019}:27-40) and the \textit{feed-forward loop motif} (\cite{Alon 2019}:41-73). They may  contain higher-dimensional interactions characterised by \textit{clique complexes} \cite{Petri  2014}. Networks
may also contain hubs, central nodes of importance \cite{Junker and Schreiber 2011}. Their nature is highly dynamic \cite{Deritel et al 2010}. 

\paragraph{Causation} Because interaction networks are \href{https://en.wikipedia.org/wiki/Graph_theory}{directed graphs}  (i.e. the edges between nodes have orientations), they represent causal effects, where causation is defined as  \cite{Pearl 2009} \cite{Pearl_Why}. \cite{Hofmeyer 2018 Constructors} shows how such diagrams can be used to exemplify
causal entailment in a diverse range of processes:  enzyme action, construction of automata,  and ribosomal polypeptide synthesis through the genetic code.

Their causal effects can be tested by \textit{experiment}, where this is possible (vary conditions at one node and show that, \textit{ceteris parabus} (i.e. conditions at other links to the node are unchanged) this results in a reliable change at another node. When this is not possible, one can use \textit{counterfactual arguments}: demonstrating that as a result of the nature of the network links this should indeed be the outcome if one were to make such a change.  This is the kind of argument I will use to claim that both upward and downward relations between levels are also causal (\textbf{Table 3}).

\paragraph{Networks and strong emergence} Because of their systemic properties, biochemical networks display strong emergence \cite{Boogerd_Networks}.

\section{Domains of Interest (DOI)}\label{sec:Domains_of_interest}

In examining the issue of causal closure of properties \textbf{P(d)}, one must have the context clearly in mind. To do so, it is useful to define the \textit{Domain of Interest} (\textbf{DOI}) of such study. This has three quite different aspects.

 First, there will a specific \textit{Topic of Interest} (\textbf{TOI}) one wishes to investigate. For example, it might be physics or engineering or chemistry or biology. In physics, one might have in mind atomic physics or condensed matter physics or plasma physics; in biology, molecular biology or physiology or neuroscience or population evolution. Or one might want to investigate relations between various of these topics. 
 
 In this paper, the interest is the nature of causal closure in the relation between physics, engineering, and biology. 

Second, given a choice of topic of interest, 
the Domain of Interest \textbf{DOI} of a system of interest $S$  consists firstly of interaction limits for $S$ with its \textit{surroundings}, and secondly of \textit{time limits} on the duration when we are interested in the behaviour of $S$. Together these comprise spacetime limitations (Section \ref{Interaction_limits}), leading to \textit{Effectively Isolated Systems} in the case of physics (Section \ref{sec:effectively isolated}) and \textit{Effective Spatial Closure} in the case of biology and engineering (Section  \ref{sec:open_controlled}).

Thirdly, there will be a choice of \textit{Levels of Interest} (\textbf{LOI}). The issue of \textbf{LOIs} is the focus of this paper, and is discussed in the following Section \ref{sec:Levels of Interest}).

\subsection{Spacetime  limitations}\label{Interaction_limits}
To be of physical interest, $S$ must be  spatially limited. Although they are often talked about, systems of infinite extent do not occur in the real universe \cite{Ellis Meissner Nicolai}.\footnote{Except perhaps for the Universe itself; but if this is indeed the case, it is of irrelevance to physics, because we can neither prove that this is the case or disprove it, and we cannot interact with or be affected by any regions outside our Particle Horizon \cite{Ellis 2014 Phil Cosm}.}  

Space time boundaries define the time and spatial domain we are interested in in relation to $S$. From a spacetime viewpoint, this is a world tube of finite radius $R$ that surrounds $S$, large enough to contain $S$ and all the elements strongly interacting with it,  bounded by an initial time $t_i$ and final time $t_f$ defined in a suitable way. This governs the kinds of interactions it can have with other systems.

\paragraph{Time limitations} We may be interested in short or long timescales characterised by the starting time $t_i$ and ending time $t_f$, depending on what we wish to study. We may be interested in,   
\begin{itemize}
	\item \textit{Evolutionary processes} \textbf{E} whereby the family of systems of similar type to $S$ came into existence over long timescales via   reproduction with variation followed by  selection; 
	\item \textit{Developmental processes} \textbf{E} whereby a specific system $S$ came into existence through developmental or manufacturing processes, or perhaps by self assembly;
	\item \textit{Functional processes} whereby the properties \textbf{P} of the system $S$  considered over short timescales emerge from the underlying physics. This is the focus of this paper.
\end{itemize}
Each involves very different choices of the timescale $\Delta t:= t_f - t_i$ relevant to our study.

\subsection{Effectively isolated systems}\label{sec:effectively isolated}
\paragraph{Isolated systems}
Causal closure of a system $S$ cannot happen if uncontrolled influences arrive from the surroundings:
\begin{equation}\label{eq:sideways}
\textbf{Sideways influences} : \{\textbf{Outside} \Rightarrow  \textbf{S} \}
\end{equation}  
As these influences vary with time, they will cause changes in the the state of the system with time that cannot be predicted from a knowledge of the properties of the system alone. The system is not causally closed.

Physics deals with this by introducing the idea of an \textit{Isolated system}. This is usually expressed by giving limits on any incoming influences ``at infinity'', for example such conditions are imposed in studying electromagnetic and gravitational radiation.

However, as just stated, infinity is not a valid physics concept.  One should instead refer to \textit{Finite Infinity} ${\cal I}$ \cite{Ellis_2002_Cosm_local}: a world tube of finite radius $R_{\cal I} >>  R$ chosen so that incoming radiation and matter  will not seriously interfere with $S$.\footnote{We cannot shield from neutrinos and gravitational radiation no matter how we choose ${\cal I}$, but this does not matter as their effects are so weak.} The dynamics of the system will then be autonomous except for small perturbations due to incoming matter and radiation crossing ${\cal I}$, which can be treated as small effects. 
\begin{quote}
	\textbf{Effectively Isolated Systems (\textbf{EIS})\textit{ What we can do is hope to find a world tube ${\cal I}$ of finite radius that serves as an effective infinity for the surface $S$. The dynamics of $S$ can be treated as an autonomous system, affected by small incoming perturbations over ${\cal I}$}.}
\end{quote}
However, there are two problems with this idea: one to do with physics, and one to do with engineering and biology.

\paragraph{Causal Domains} The first is that famously, in general relativity, causal domains are determined by null cones rather than timelike tubes \cite{Hawking and Ellis 1973}
. Why have I not defined the causal limits, which are basic in term of causal closure, in terms of null surfaces rather than a timelike world tube?

The answer is that on astronomical scales, effective causal limits are indeed given by timelike world tubes rather than null surfaces. On astronomical scales at recent times, the dynamic effects of radiation are very small compared with those of matter. We get a very usable ${\cal I}$ by choosing $R_{\cal I}$ to be about 1 Megaparsec in comoving coordinates \cite{Ellis and Stoeger 2009}. Nothing outside there has had a significant effect on the history of our galaxy or the Solar System. Yes some radiation and matter is coming in, but it is negligible compared to the energies involved in daily life. The one form of radiation of significance for the world is light from the Sun, which comes from well within those limits: 1 Astronomical Unit might indeed suffice for local physics. The radiative energy coming from greater distances has negligible dynamical effects on Earth. 
 A  timelike world tube of radius 1AU will do just fine in terms of considering causal closure of the Solar System.

\paragraph{Isolated systems: Laboratories}
However physics practice works in a different way: the key concept is an isolated system in a laboratory.  It's a system that is in fact interacting strongly with the the environment (Section \ref{sec:open_controlled}), but that interaction is strictly controlled so that it is highly predictable. The system is shielded from influences outside the laboratory as far as possible. 
This then enables the results of experiments to also be highly predictable. And that is what enables the determination of the Effective Theories $\textbf{ET}_\textbf{L}$ that hold at a Level \textbf{L}. Examples are the expensive isolation and cooling systems underlying the success of
\href{https://www.sciencedirect.com/topics/physics-and-astronomy/quantum-optics}{quantum optics}  experiments.
  
\paragraph{Engineering and Biology as Open Systems}
The real problem is different. It is that no biological system can be closed: they have by their nature to be open systems.  And the same is true for engineering systems. In these cases, the `isolated system' paradigm is simply wrong.

\subsection{Effective Spatial Closure}\label{sec:open_controlled}
Life cannot exist as an isolated system. \href{https://en.wikipedia.org/wiki/Living_systems}{Biological systems}  are inherently open systems interacting with the environment \cite{Hoffman}.  (\cite{Peacocke 1990}:10-11) states, 
\begin{quote}
	``\textit{Biological organisations can only maintain themselves in existence if there is a flow of energy and this flow requires that the system not be in equilibrium and therefore spatially inhomogeneous''.}\footnote{This is a kind of symmetry breaking different than the cases discussed in \cite{Ellis_2020_response}, and therefore may, by the same kind of arguments,  be related to strong emergence in biology}.
\end{quote} 
The effect of the outside world is not negligible. On the contrary, it is essential to biological functioning. It cannot be treated as a perturbation. The \href{https://en.wikipedia.org/wiki/Biosphere}{biosphere} experiences incoming high grade radiation from the Sun and radiates outgoing low grade heat to the dark Sky, and this is its energy source enabling it to function thermodynamically \cite{Penrose 1990 MInd}. Organisms need a flow of material in, and, because of the \href{https://en.wikipedia.org/wiki/Second_law_of_thermodynamics}{Second Law of Thermodynamics} (essentially: as time progresses, matter and energy will be transformed from usable to unusable forms), need to dispose of waste matter and heat resulting from internal non-equilibrium metabolic processes. A living system \textbf{S} must take in materials and energy from the surrounding environment \textbf{E} and 
dispose of waste matter and energy to \textbf{E}:
\begin{equation}\label{key}
\textbf{E} \rightarrow \textbf{S} \rightarrow \textbf{E}.
\end{equation}
In summary, living systems are essentially interacting systems. The same is true for engineering systems, because they do work of some kind. 
   
\paragraph{Reliable Interactions} They must therefore interact strongly with an environment that is stable enough that  
the interactions with the environment are reliable and reasonably constant so they do not disrupt the dynamics of the system over time. 

In physics this is the concept of a Heath Bath or \href{https://en.wikipedia.org/wiki/Thermal_reservoir}{Thermal reservoir}. You are in contact with an environment but don't need to take its dynamics into account because it is in a state usually assumed to be static, characterised only by a constant temperature $T$; and it is so large that the system $S$ has negligible influence on it state.\footnote{Actually the  assumption that $\partial T/\partial t = 0$ is problematic if one wants to explain the arrow of time \cite{Ellis_2020_response}; but I'll leave that aside for the present.}

\paragraph{Life: Interaction limits} 

How then does one limit those interactions to those that will enable life to sustain itself? Humberto Maturana  and Francisco Varela essentially dealt with this by introducing the idea of \href{https://en.wikipedia.org/wiki/Autopoiesis}{\textit{Autopoesis}} \cite{Maturana and Varela 1980}, which \textit{inter alia} expresses the idea of causal closure in terms of system interaction with its environment.

\paragraph{System Boundaries}
The key point is how does one define the boundary of a system in this context. Instead of choosing a spacetime tube of some chosen radius $R$ as in the astronomy case, one chooses a \textit{System Boundary} \textbf{B} that characterizes the system as being effectively autonomous.  This underlies the meaning of Level \textbf{L8} in the hierarchy of emergence (\textbf{Table 1}, \S \ref{sec:hierarchy_nature}). A person has a \href{https://en.wikipedia.org/wiki/Skin}{skin} that is her physical boundary with the outside world, but still allows interaction with it. Energy and matter transfer takes place across the boundary. A \href{https://en.wikipedia.org/wiki/Machine}{machine} similarly has a boundary that defines its limits, but will have some form of energy input enabling it to do work, and in many cases complex cooling devices to get waste heat out (paradoxically, they may consume large amounts of energy).
 
\noindent But this idea extends down to other levels, for example it holds also at the Device Level/Cellular Level \textbf{L5} and the Component Level/Organ Level \textbf{L6}. For example, a \href{https://en.wikipedia.org/wiki/Cell_biology}{cell} is the core of biology. It exists as an entity with its own integrity, characterised by the \href{https://en.wikipedia.org/wiki/Cell_wall}{cell wall} which allows controlled ingress and egress of materials and energy, yet interacts strongly with the environment.    \\

\noindent How then does life handle functioning in this context? The environment must be sufficiently stable so as to allow effective predictability.
This is the case when the system is not causally closed but has predictable interactions with the environment that makes its own functions  predictable. It has an environment that can be treated as predictable up to perturbations. The environment may change with time, but if so, slowly enough to allow adaptation to the changing situation.
This is often the case, and is what on the one hand allows living systems to flourish, and on the other allows biology to be a genuine scientific subject \cite{Campbell and Reece 2005}.

\paragraph{The exceptions}
But that is not always the case: take COVID-19 as an example. That started at the social level (\textbf{L9}) in one house, then spread worldwide via aircraft affecting life world wide and thereby affecting in a downward way all the biomolecules (\textbf{L4}) and electrons (\textbf{L1}) in the bodies of doctors and nurses and patients effected. But additionally it affected the engineering side by closing down thousands of flights across the world, thereby reaching down to affect all the billions of atoms (\textbf{L3}) and particles (\textbf{L1}) that make up those aircraft: a classic example of the interlevel causation I turn to next. 

As far as predictability is concerned, the system $S$ has reliable  interactions with the environment and so is predictable most of the time, except when we this is not the case. Predictability holds when all things are equal, but there is no guarantee this will be the situation. And that is the best we can do.\footnote{And no this could not be predicted in principle in a strictly bottom up way if we knew the detailed  positions and momenta of all the particles in our causal past to utmost precision, for multiple reasons \cite{Ellis_2020_response} \cite{Ellis 2020a}. These are strongly emergent phenomena, unpredictable even in principle.}
\begin{quote}
	\textbf{Effective Spatial Closure (ESC)  \textit{Engineering and biological systems of necessity interact strongly with their environments, because that is necessary for their functioning. One can in practice usually set up a situation of} Effective Spatial Closure \textit{where that interaction is by and large predictable so that the system will act in a predictable way. However there is no guarantee that this effective causal closure will always be as respected by the environment. Effective Spatial Closure is largely reliable, but  holds} ceteris parabus}. 
\end{quote}	
	Elaborating, \href{https://www.investopedia.com/terms/c/ceterisparibus.asp}{Investopaedia} states this concept as follows:
\begin{quote}
	``Ceteris paribus \textit{ is a Latin phrase that generally means `all other things being equal.' In economics, it acts as a shorthand indication of the effect one economic variable has on another, provided all other variables remain the same. ... Ceteris paribus assumptions help transform an otherwise deductive social science into a methodologically positive `hard' science. It creates an imaginary system of rules and conditions from which economists can pursue a specific end. Put another way; it helps the economist circumvent human nature and the problems of limited knowledge'}.
\end{quote} 
In other words, Effective Spatial Closure works except when it doesn't. Unpredictability happens when  \href{https://en.wikipedia.org/wiki/Black_swan_theory}{\textit{Black Swan Events}} take place, possible when there is a fat-tailed  rather than a Gaussian distribution \cite{Taleb 2010}, and with major significance at the macro level. 
And when it doesn't work, the effects chain all the way down from Level \textbf{L8} to the atomic level \textbf{L3} and particle level \textbf{L1} in \textbf{Table 1}, as in the case of the Cornovirus pandemic. 

\section{Levels  of Interest (\textbf{LOI})}\label{sec:Levels of Interest}
Setting this aside, the key conceptual issue I will deal with in this paper is the relation of the  causal closure of emergent properties \textbf{P} to levels 
 in the hierarchy of emergence. 

Effective Theories and Existence of Levels is discussed in Section \ref{sec:complete_interactions}. Levels of Interest (\textbf{LOIs}) are defined in Section \ref{sec:LOI_def}. Sensible  choices for  LOIs are discussed in Section \ref{sec:LOI_choice}.
However there is a problem: Interactions span all levels: every LOI interacts with every other level by both upwards and downwards causation,  so how can one get  meaningful \textbf{LOIs}, or indeed a meaningful level? I elaborate on this problem in Section \ref{sec:LOI_restricted}.   A practical way out is by defining Effective Causal Closure, discussed later in  Section \ref{sec:causal_closure_biology}.  
The choices one makes relate to whether one wants to answer \textit{How} questions or \textit{Why} questions (Section \ref{sec:LOI_how_and_why}).   

\subsection{Effective Theories and Existence of Levels}\label{sec:complete_interactions}

\textbf{An Effective Theory}
$\textbf{ET}_\textbf{L}$ (Section \ref{sec:equal_valid}) is a set of variables and equations  representing interactions and constraints at a particular level \textbf{L}, such that 
 initial data implies a reliable outcome at that level. 
It is the possibility of existence of 
an Effective Theory $\textbf{ET}_\textbf{L}$ (\ref{eq:effective_laws}) 
   at each Level \textbf{L}  (the dynamics at that level is determined at that level) 
 that underlies the very 
  concept of levels in the first place. 
   As commented by Arthur Peacocke (\cite{Peacocke 1990}:10), following from \cite{Simon 2019}
\begin{quote}
	\textit{``Natural hierarchies are more often than not `nearly decomposable' - that is, the interactions among the sub-systems (the `parts') are relatively weak compared withe the interactions among the subsystems, a property which simplifies there behaviour and description''}
\end{quote}
 The fact that such levels 
  exist is a consequence of the nature of the underlying physical laws and the values of the constants of nature \cite{Uzan}. It allows the existence of the modular hierarchical structures that are the core foundation of complexity (\S \ref{sec:hierarchy_nature}, \S \ref{sec:modular}). 
\subsection{Definition of Levels of Interest \textbf{LOI}s}\label{sec:LOI_def} 
\cite{Simon 2019}) defines a hierarchical system as follows, quoted in (\cite{Peacocke 1990}:249):
\begin{quote}
	 ```\textit{\textbf{A hierarchical system}: A system of composed of inter-related subsystems, each of the latter being in turn, hierarchical in structure until we reach some lowest level of elementary subsystem  (the choice of this lowest level he regarded as arbitrary)}''.
\end{quote}
This is what I have taken for granted above. While it has all the levels shown in \textbf{Table 1} (\S \ref{sec:hierarchy_nature}), we usually do not want to consider them all at once. \cite{Blundell 2019} says it thus:
\begin{quote}
	``\textit{Thus we take it as a given that when a portion of the universe is selected for study, be it a gas	or a galaxy, we are allowed to blissfully ignore what is going on at scales that are much larger, or indeed much smaller, than the one we are considering.} ''
\end{quote}
That is what I am formalizing by defining the concept of Levels of Interest (\textbf{LOI}). They are defined as follows: 
\begin{quote}
	\textbf{Levels of Interest} (\textbf{LOIs}) \textit{\textbf{ is a definition of the range of levels that will be covered by a theory
	}}
\end{quote}
A \textbf{LOI} is defined by its top level \textbf{TL} and its bottom level \textbf{BL}, thus
\begin{equation}\label{key}
\textrm{\textbf{LOI(TL-BL)}} := \{\textbf{TL} \cup  \textbf{BL}\}
\end{equation}  
where the Union sign $``\cup''$
 means include all levels between \textbf{TL} and \textbf{BL}. 
Some studies are unilevel: \textbf{TL = BL}, and some are explicitly interlevel: $\textbf{TL} > \textbf{BL}$.
 
\begin{itemize}
	\item One can validly define such a LOI  regardless of what levels you choose, because the levels are equally causally valid (Section \ref{sec:equal_valid}). 
	\item Your choice will depend on your Topic of Interest (\textbf{TOI}) (see \S \ref{sec:Domains_of_interest}). It is helpful if the choice of levels is made explicit, e.g. \textbf{LOI(3-5)} covers levels 3 to 5
	\item Then one can for the purposes of studying  the dynamics at those levels legitimately ignore higher and lower levels, in the sense to be explored below.
	\item This is possible because the levels effectively decouple in the sense of allowing valid Effective Theories $\textbf{ET}_\textbf{L}$ at each level \textbf{L}.
	\item However this does not give you the right to deny the validity of levels that lie outside your Levels of Interest.
	\item There is no guarantee that causal completion will occur by including only those levels characterised by your choice of \textbf{LOI}.
\end{itemize}
That is the topic of Interlevel Causal Closure, which  I discuss below (Section \ref{sec:interlevel closure}).

\subsection{Choice of \textbf{LOI}s}\label{sec:LOI_choice}
Are there limits on the \textbf{LOI}s one can choose? 
One can choose to investigate any desired
LOIs, not investigating or take for granted the
interactions that will inevitably occur from
higher and lower levels. This is done to
establish $\textbf{ET}_\textbf{L}$s.

For example Denis Noble in \cite{Noble 2002} ``\textit{Modeling the heart--from genes to cells to the whole organ}'' chooses to investigate the range of levels stated, namely Levels \textbf{L5} to \textbf{L8} in the Hierarchy of Emergence (\textbf{Table 1}),  and not for example on the one hand the interactions between protons and electrons that make this possible, and on the other hand the mental and social influences that will inevitably be having an effect on how the heart is functioning. What he does do is investigate the interlevel relations within the levels \textbf{{LOI(5-8)}} he has chosen, because that is the domain of physiology.

 Other examples are shown in \textbf{Table 2}.\\

 
 \begin{tabular}{|l|l|}
 	\hline 
Particle Physics \cite{Oerter 2005} &  \textbf{L1} \\ 
Nuclear Physics	(\cite{Hewitt 2015} \S 33-34) & \textbf{L1}-\textbf{L2} \\ 
Atomic Physics (\cite{Hewitt 2015} \S 32)	&  \textbf{L2}-\textbf{L3}\\
Solid State Physics \cite{Simon 2013} &  \textbf{L3}-\textbf{L4}\\
Computer Structure  \cite{Tannenbaum} & \textbf{L5}-\textbf{L7}\\
The Molecular Biology of the Gene \cite{Watson et al 13} & \textbf{L3}-\textbf{L4}\\
 The Molecular Biology of the Cell \cite{Alberts_etal_07} &\textbf{L3}-\textbf{L5}\\
 Neuroscience  \cite{Kanetal13} &\textbf{L3}-\textbf{L6}\\
 Physiology \cite{Physiology_Animal} \cite{Physiology_human}& \textbf{L4}-\textbf{L7}\\
 Biology \cite{Campbell and Reece 2005} & \textbf{L4}-\textbf{L8}\\
 Major Transitions in Evolution \cite{Maynard Smith transitions} 
& \textbf{L4}-\textbf{L9} \\
Global Climate Change \cite{Houghton 2009} &\textbf{L3}-\textbf{L9}\\
\hline 
\end{tabular} \\

\textbf{Table 2:} \textit{Levels of Interest} \textbf{LOIs} \textit{for various academic disciplines}.\\

\noindent  
Physics right down to Level \textbf{L1} always underlies what is happening, even if it lies outside the levels of interest to you. What happens in the real world  right down to the physics levels \textbf{L1-L3} is always influenced by what happens at higher levels including \textbf{L9}, even if that is not what interests you in your particular studies.   

	\noindent What is a sensible DOI depends on conditions.
\begin{quote}
\textbf{Choice of Levels of Interest} \textit{It all depends on what you want to understand. The suitable choice of \textbf{LOIs} will follow. This is the context within which ``causal completeness of physics'' must be evaluated}.
\end{quote}

\subsection{Interactions span all levels}\label{sec:LOI_restricted}
\paragraph{Strongly Interacting Levels}
There is a basic problem with LOIs, however. That is the fact that, as mentioned above,  every level interacts with every other level! 
The choices one makes relates \textit{inter alia} as to whether one wants to answer \textit{How} questions, \textit{Why} questions, or both (Section \ref{sec:LOI_how_and_why}).  There is a practical way out that I discuss in Section \ref{sec:causal_closure_biology} by introducing the idea of \textit{Effective Causal Closure}.

\subsection*{The Whole Shebang}\label{sec:LOI_whole}
\textbf{Table 1} omits two levels. At the bottom, it omits Level \textbf{L}: the level of a \textit{Theory of Everything} (\textbf{TOE}). 
At the top, it omits Level \textbf{L10}: the level of Cosmology. 

\paragraph{At the bottom: The TOE and dynamic properties P(d)}

In order to examine the emergence of 
 machines on the engineering side and organisms on the biology side, all one needs is Newton's Laws of motion, Maxwell's equations, Galilean gravity, and maybe the Schr\"{o}dinger equation (\cite{Laughlin_Pines_2000}; \S 4.1 in \cite{Ellis_2020_response}).  That is, Level \textbf{L3} is an adequate base. All engineering and biology emerges from this level,
it is the lowest level engineers and biologists need to study. As explained in \cite{Ellis_2020_response} and \S\ref{sec:upward_downward}, conversely engineering  and biology reach down to shape outcomes at Level \textbf{L3} via time dependent potentials or constraints.

However Levels \textbf{L3} and \textbf{L2} emerge from the  Quantum Field Theory and the Standard Model of Particle Physics at Level \textbf{L1}  \cite{Oerter 2005}. Thus  \textbf{L1} is a deeper foundation of  emergence. However it is essentially decoupled from everyday life \cite{Anderson_72} \cite{Laughlin_Pines_2000}. Nevertheless  outcomes at Level \textbf{L1} too must also be  shaped in a downward way by engineering and biological variables, via outcomes at Level \textbf{L3}. 

But this is not the bottom. Underlying \textbf{L1} is some theory of fundamental physics at Level \textbf{L0}, a ``Theory of Everything'' (\textbf{TOE}) \cite{Weinberg 1994}: maybe String Theory/M Theory, maybe not. There are variety of competing theories on offer 
\cite{Murugan Weltmann Ellis 2012}. This is the ultimate source of the emergence of dynamic properties \textbf{P}(\textbf{d}) in engineering and biology (using the notation in \S2.2 of \cite{Ellis_2020_response})). It affects us every day.

These lie outside the \textbf{LOIs} of engineers and biologists. That is just as well, as we do not know what the answer is at Level \textbf{L0}, even though it underlies all physical emergence. As explained above, it suffices to deal with Effective Theories that hold at higher levels.

\paragraph{At the top: Cosmology and Existence E}
Consider Isaac Newton seeing an apple drop. This occurs for a variety of reasons: the Law of Gravity acting on the apple, the light rays that convey this image to his retina, the analysis of the image by his brain, and so on. But there are far deeper underlying issues. Why does the apple exist? Why does the Earth exist? Why does the Solar System and the Galaxy exist? Why does the Universe exist and have the nature it does? These are all the background reasons why the apple fell, and why Isaac Newton existed for that matter. 

These questions of how everything came into being  (\textbf{E}) is the domain of Cosmology \cite{Peter and Uzan 2013} . It deals with issues such as, Where do elements such as Helium and Carbon come from? How did the Galaxy, the Sun, and the Earth arise? The Philosophy of Cosmology \cite{Ellis 2014 Phil Cosm} considers issues such as Why is physics of such a nature as to allow life to exist? \cite{Sloan} Particularly: why are the constants of nature \cite{Uzan} of such a character as to allow the hierarchical structure in \textbf{Table 1} to emerge? 
Thus cosmology affects us every day by underlying our existence.

\paragraph{Everyday effects of cosmology P}
There are more immediate issues as well, in the relation of cosmology to everyday life \cite{Ellis_2002_Cosm_local}: Why is the Sky dark at night, serving as a heat sink for the Earth's waste energy? This is crucial to the functioning of the biosphere (\cite{Penrose 1990 MInd}:411-417).  Why is there an arrow of time? \cite{Penrose 1990 MInd} \cite{Davies 2004}
 \cite{Ellis_Drossel_time}? This is crucial to all macro level physics, biology, and chemistry. Both are due to the cosmological context. Thus cosmology affects the emergence of properties \textbf{P} in engineering and biology today.

The point then is that while it does indeed have a major causal effect on daily life \cite{Sciama 2012}, this is a rock solid relation that does not change with time. It is a fixed unchanging background that does not alter effective laws as time passes. It is thus not a case of \textit{ceteris parabus}\footnote{Unless we live in a \href{https://en.wikipedia.org/wiki/False_vacuum}{false vacuum} in which case local physics could suddenly change in a way that might wipe out all life \cite{Tegmark 2005}.} (c.f.\S\ref{sec:open_controlled}\') and so can be taken for granted and not considered further when investigating causal closure  in engineering, biology, and physics.

\subsection{How questions and Why questions}\label{sec:LOI_how_and_why}
An issue in choosing \textbf{LOIs} is if one is  interested in \textit{How} questions or \textit{Why} questions.\\

\textbf{How questions} consider physical interactions on the one hand, and mechanisms on the other. Thus they will relate to levels \textbf{L1}-\textbf{L7} in \textbf{Table 1}, including \textbf{L7} because that relates to the integration of systems to produce the organism as a whole \cite{Physiology_Animal} \cite{Physiology_human}.\\

\textbf{Why questions} relate to motivation, meaning, and philosophical issues. Thus they will relate to levels \textbf{L7}-\textbf{L8} in \textbf{Table 1}, including \textbf{L7} because this is the level where as well as philosophy, psychology and motivation come in  \cite{Donald 2001}  \cite{Kandel 2012} and \textbf{L8} where the causal power of social structures enters  \cite{Vass}. 

There is of course a trend for some strong reductionists to deny that the Why questions are valid or have any real meaning. From the viewpoint of this paper, that simply means that they themselves have a restricted set of \textbf{LOIs} that excludes those higher levels. Because of the equal validity of levels espoused in \cite{Noble 2012}, \cite{Ellis_2020_response}, and in this paper, that restricted set of interests does not provide a justification for denying the validity of the levels with Effective Theories outside their particular set of interests. 

\section{Interlevel Causal Closure: Biology}\label{sec:interlevel closure}
The crucial concept in this paper is that of \textbf{Interlevel Causal Closure} of properties \textbf{P(d)}. I first  consider Causation and Causal Closure (Section \ref{sec:causation}) and the nature of biology (Section \ref{sec:Nature_of_biology}). 
Then I consider Interlevel Causal Closure in the case of biology (Section \ref{sec:causal_closure_biology}), because this is where it is clearest and has been discussed most. 
I introduce here the key concept of \textbf{\textit{Effective Causal Closure}}. A stronger relation is the idea of  \textbf{\textit{Inextricably Intertwined Levels} }(\textbf{IIL}s) which I discuss in Section \ref{sec:Intertwined_Biology}. 
\subsection{Causation and Causal Closure}\label{sec:causation}
\paragraph{Causation} In order to consider causal closure, one must first have a view on how one justifies claims of causation. This has been laid out in depth by Judea Pearl in \cite{Pearl 2009}  and \cite{Pearl_Why}.  Causal inference is based in  Causal Models (directed graphs) validated by experimental intervention, or when that is not possible, by  Counterfactual Arguments. 

\paragraph{Causal Models}
Here one consider causal models of the influences at work .
In effect, the diagram of the hierarchy of Levels  Table 1 in Section \ref{sec:hierarchy} \textit{is} such a (very simplified) model, when one introduces the arrows of both upward emergence (left) and downward constraint or control (right).

\vspace{0.1in}
	\begin{tabular}{|c|c|c|c|c|c|}
	\hline \hline
	& \textbf{BU} &\textbf{Engineering} & \textbf{Life Sciences} & \textbf{TD}\\	
	\hline Level 9 (\textbf{L9})& & Environment &  Environment &$\Downarrow$ \\ 
	\hline 
	Level 8	(\textbf{L8})&$\Uparrow$ & Sociology/Economics/Politics &  Sociology/Economics/Politics & $\Downarrow$ \\ 
	\hline 
	Level 7	(\textbf{L7})&$\Uparrow$ & Machines &  Individuals & $\Downarrow$\\ 
	\hline 
	Level 6	(\textbf{L6})&$\Uparrow$ & Components &  Organs & $\Downarrow$\\ 
	\hline 
	Level 5	(\textbf{L5})& $\Uparrow$& Devices & Cells  &$\Downarrow$ \\ 
	\hline 
	Level 4 (\textbf{L4})&$\Uparrow$	& Crystals & Biomolecules & $\Downarrow$\\ 
	\hline 
	Level 3 (\textbf{L3})&$\Uparrow$ & Atomic Physics	& Atomic Physics & $\Downarrow$
	\\ 
	\hline 
	Level 2	(\textbf{L2})& $\Uparrow$& Nuclear Physics & Nuclear Physics  & $\Downarrow$\\ 
	\hline 
	Level 1	(\textbf{L1}) & $\Uparrow$ &  Particle Physics & Particle Physics  & \\ 
	\hline \hline
\end{tabular} 
\vspace{0.1in}

\textbf{Table 3}: 
\textit{The emergent hierarchy of structure and causation for engineering  (left) and life sciences (right), indicating the upward and downward causation occurring.}\\

\noindent To develop this approach more fully, one needs to expand \textbf{Table 3} to a hierarchical diagram that  represents the modular nature of the hierarchy (\S\ref{sec:modular}). This is a very worthwhile project, but I will not attempt it here. It is roughly indicated in (\cite{Peacocke 1990}:8-11,247-248), and examples are in (\cite{Buchanan 2010}:6,10-11,22,95,110,132,160). 

\paragraph{Intervention} Here one actually intervenes at Level \textbf{LI} and reliably observes a resultant change at level \textbf{LF}. This has been done both for Effective Theories $\textbf{EF}_\textbf{L}$ at each level \textbf{L}, and in many case for both upwards and downwards interlevel effects. One can do this also using digital computer models; for example  \cite{Fink_and_Noble} \cite{Noble 2012} have done this to show downwards causation occurring in  computer models of heart function.

\paragraph{Counterfactual views} \cite{counterfactual} Here one considers what would happen if one intervened at Level \textbf{LI}, and plausibly argues that this will cause an actual difference at Level \textbf{LF}, when upward causation takes place:  $\textbf{LF} > \textbf{LI}$ (left column \textbf{BU}) or when downward causation takes place: $\textbf{LI} > \textbf{LF}$ (right column \textbf{TD}). This has been used to establish that downward causal effects exist, e.g. \cite{Campbell_1974}, \cite{Ellis_Kopel_2019}, \cite{Ellis_Drossel_computers}.

\paragraph{Causal Closure}\label{sec:causal closure}
Consider a multilevel system \textbf{S} (which could have only one level).
\begin{quote}
	\textbf{ Causal Closure of the properties \textbf{P} of a system $\textbf{S}(\textbf{BL-TL})$ with a bottom level \textbf{BL} and top level \textbf{TL} \textit{ occurs when 
	the set of Effective Laws $\textbf{EF}_\textbf{L}$ governing outcomes at each level \textbf{L}, together with the upward and downward interactions between levels, are sufficient to determine the future state  $\textbf{o}$ of the system $S$ at all levels \{\textbf{BL-TL}\} (the outcome) 
			from  an initial state $\textbf{d}$ (the data) given at a set of levels \{\textbf{LL-HL}\} with lowest level \textbf{LL} and highest level \textbf{HL} contained within or equal to \{\textbf{BL-TL}\}.}
}\end{quote}
Note that this includes the physicalist idea of causal closure where all follows from a single lowest physical level \textbf{LL}, for that is the case $\textbf{LL = HL}$, chosen as \textbf{L1} in \textbf{Table 1}. \\

\noindent  Causal Closure requires Effective Predictability on the one hand (Section \ref{sec:open_controlled}), and
 an Effectively Causally Closed  set of levels on the other, which concept I now consider. 
 
 \paragraph{The issue: Two opposing strands}
 \begin{itemize}
 	\item There are no isolated sets of levels, as just discussed (Section \ref{sec:LOI_restricted}). Causal Closure as just defined is an ideal that does not occur in practice unless one takes \textbf{LL = L1}, \textbf{HL = L9}: you are giving data for the whole thing. 
 \item However there are in practice preferred restricted sets levels with a special integrity  in terms of causal closure (see the comments just after \textbf{Table 1}).
 \end{itemize}
How do we deal with this tension? The clearest domain in which to tackle this is biology (\S \ref{sec:causal_closure_biology}). The lessons from there carry over to engineering (\S \ref{sec:Effective_computer}), and physics (\S \ref{sec:Intertwined_phyiss}).
\subsection{The nature of biology}\label{sec:Nature_of_biology}
Biological organisms have purpose, as stated by Nobel Prize winning biologist Leland Hartwell and colleagues  \cite{Hartwell_et_al_1999}:\footnote{And see also \cite{Moss and Nicholson 2012}.} 
\begin{quote}\textit{``Although living systems obey the laws of physics and chemistry, the notion of function or purpose differentiates biology from other natural sciences. Organisms exist to reproduce, whereas, outside religious belief, rocks and stars have no purpose. Selection for function has produced the living cell, with a unique set of properties that distinguish it from inanimate systems of interacting molecules. Cells exist far from thermal equilibrium by harvesting energy from their environment. They are composed of thousands of different types of molecule. They contain information for their survival and reproduction, in the form of their DNA. Their interactions with the environment depend in a byzantine fashion on this information, and the information and the machinery that interprets it are replicated by reproducing the cell.}''
\end{quote}
Consequently, as emphasized by Peacock (\cite{Peacocke 1990}:13)
\begin{quote}
	``\textit{Many biological concepts and language are often} sui generis \textit{and not reducible to physics and chemistry, certainly not in the form to which they apply to simpler and restricted atomic and molecular  systems}''. 
\end{quote}
In the case of biology, unless the concepts considered include purpose and function, it will miss the essence of what is going on, as pointed out by \cite{Hartwell_et_al_1999}. You also need to introduce the concepts ``alive'' and ``dead'', which do not occur at any lower level than the cellular level in biology, and do not occur at any physics level. Without this concept you cannot for example discuss the theory of natural selection \cite{Mayr 2001}.

\paragraph{Upward and downward causation} 
 As just stated, all biological entities have purpose or function, and that controls in a top-down way what happens at lower levels \cite{Noble 2012} reaching down to the underlying physical levels \cite{Ellis_Kopel_2019}. The physics does not control the higher levels, rather - without any violation of the laws of physics - it does what the biology asks it do. 
 This functioning occurs via a combinations of upwards and downwards causation \cite{Noble2008_Music} \cite{Noble 2012}, for example gene regulation taking place on the basis of the state of the heart \cite{Fink_and_Noble} or the brain \cite{Kandel_memory}. 
 This dynamic reaches down to the molecular level and then the underlying  electron level. 
 
 The enabling factors are \href{https://en.wikipedia.org/wiki/Blackboxing}{black boxing} \cite{Ashby 2013} to get higher level logic out of lower level logic, together with  time dependent constraints at the lower level that are regulated by higher level biological variables   
 \cite{Ellis_Kopel_2019}.
 Together they underlie the emergent effective laws $\textbf{ET}_\textbf{L}$ at each level \textbf{L} in biology.

\paragraph{Preferred levels}
The cellular level \textbf{L5} is a key level in biology: cells have an organisation and integrity of their own, and are living integral entities that are the basic units of life. They interact with other cells at the same level, and react to their environment in appropriate ways. In multicellular organisms they depend on higher levels for nutrition, materials, waste disposal, and signals as to what to do. But they can be treated as modules (\S \ref{sec:modular}) with an integrity of their own that responds to inputs and produces outputs. The associated set of biological levels, taking the underlying physics for granted, is the set of levels \textbf{L4-L5}.

 Similarly the level of individual organisms \textbf{L7} again represents a level of emergent\textbf{} integrity. Individuals are entities that can be treated as autonomous entities that respond to environmental cues (from the levels above) and other individuals (at the same level). The associated set of biological levels, taking the underlying physics for granted, is the set of levels \textbf{L4-L7}.
 
 So the issue is, how does this kind of autonomy emerge at these particular levels, given that all levels are interacting?
 
\subsection{Effective  Causal Closure in Biology}\label{sec:causal_closure_biology}
 
What characterizes these special sets of levels?
The key point as to what occurs 
is  organisational closure in biological organisms 
\cite{Mosiooetal_function} \cite{Mossio  and  Moreno 2010}:
\begin{quote}\textit{
		``The central aim of this paper consists in arguing that biological organisms
		realize a specific kind of causal regime that we call `organisational closure'; i.e., a distinct
		level of causation, operating in addition to physical laws, generated by the action of
		material structures acting as constraints. We argue that organisational closure constitutes
		a fundamental property of biological systems since even its minimal instances are likely to
		possess at least some of the typical features of biological organisation as exhibited by more
		complex organisms.'' 
	}
\end{quote}
This is a distinct causal regime, as explained in  \cite{Mossio 2013}: 
\begin{quote}
\textit{``In biological systems, closure refers to a holistic feature such that their constitutive processes, operations and transformations (1) depend on each other for their production and maintenance and (2) collectively contribute to determine the conditions at which the whole organization can exist. According to several theoretical biologists, the concept of closure captures one of the central features of biological organization since it constitutes, as well as evolution by natural selection, an emergent and distinctively biological causal regime.''}
\end{quote}
This is developed further in \cite{Montevil et al constraints}, identifying biological organisation as closure of constraints
\begin{quote}
	\textit{``We propose a conceptual and formal characterisation of biological organisation as a
	closure of constraints. We first establish a distinction between two causal regimes at work
	in biological systems: processes, which refer to the whole set of changes occurring in non-equilibrium
	open thermodynamic conditions; and constraints, those entities which, while
	acting upon the processes, exhibit some form of conservation (symmetry) at the relevant
	time scales. We then argue that, in biological systems, constraints realise closure, i.e.
	mutual dependence such that they both depend on and contribute to maintaining each
	other.''}
\end{quote}
Thus biological organisation is an interlevel affair, involving downward causation as well as upwards emergence, thus enabling teleology \cite{Mossio and Bich }, \cite{Bich_2020}. From the viewpoint of this paper, these authors are identifying  specific sets of levels where effective interlevel causal closure occurs: the topmost level links to the bottom-most level to close the dynamic loop that leads to biological emergence.

 I will quote three more papers that have essentially the same view.
\cite{Hofmeyer 2017 Anticipation} emphasizes this property in the case of the cell:
\begin{quote}
\textit{
	[The] property of self-fabrication is the most basic expression
	of biological anticipation and of life itself. Self-fabricating systems must be
	closed to efficient causation... I identify the classes of efficient
	biochemical causes in the cell and show how they are organized in a hierarchical
	cycle, the hallmark of a system closed to efficient causation. Broadly speaking, the
	three classes of efficient causes are the enzyme catalysts of covalent metabolic
	chemistry, the intracellular milieu that drives the supramolecular processes of
	chaperone-assisted folding and self-assembly of polypeptides and nucleic acids
	into functional catalysts and transporters, and the membrane transporters that
	maintain the intracellular milieu, in particular its electrolyte composition.
}\end{quote}
You need all these components and levels for the thing to work.
\cite{Farnsworth 2018} emphasizes that multi-level homeostasis is part of the mix:
\begin{quote}
	\textit{Two broad features are jointly necessary for autonomous agency: organisational closure and
	the embodiment of an objective-function providing a `goal': so far only organisms demonstrate both.
	Organisational closure has been studied (mostly in abstract), especially as cell autopoiesis and the
	cybernetic principles of autonomy, but the role of an internalised `goal' and how it is instantiated by
	cell signalling and the functioning of nervous systems has received less attention. Here I add some
	biological `flesh' to the cybernetic theory and trace the evolutionary development of step-changes
	in autonomy: (1) homeostasis of organisationally closed systems; (2) perception-action systems;
	(3) action selection systems; (4) cognitive systems; (5) memory supporting a self-model able to
	anticipate and evaluate actions and consequences. Each stage is characterised by the number of
	nested goal-directed control-loops embodied by the organism, summarised as will-nestedness.} 
\end{quote}

Finally \cite{Noble and Noble 2019 A-Mergence}  argue for circular causality:
\begin{quote}
	``\textit{We argue that (1) emergent phenomena are real and important; (2) for many of these, causality
	in their development and maintenance is necessarily circular; (3) the circularity occurs between
	levels of organization; (4) although the forms of causation can be different at different levels,
	there is no privileged level of causation a priori: the forms and roles of causation are open to
	experimental investigation; (5) the upward and downward forms of causation do not occur in
	sequence, they occur in parallel (i.e. simultaneously); (6) there is therefore no privileged direction
	of emergence - the upper levels constrain the events at the lower levels just as much as the
	lower levels are necessary for those upper-level constraints to exist. 
	Modern biology has confirmed [...]
		that organisms harness stochasticity at low levels to generate their functionality. This
	example shows in fine detail why higher-level causality can, in many cases, be seen to be more
	important than lower-level processes.''
}\end{quote}
This is closely related to the idea of \textit{automous systems}, characterized by their organizational and operational
closure \cite{Villalobos and Dewhurst 2018}, where 
\cite{Thompson 2007} 
\begin{quote}
``\textit{\textbf{Organizational closure} refers to the self-referential (circular and
	recursive) network of relations that defines the system as a unity, and \textbf{operational closure}
	to the re-entrant and recurrent dynamics of such a system}
\end{quote}
which is just the idea above. This discussion is related to the idea of 
\href{https://en.wikipedia.org/wiki/Autopoiesis}{Autopoiesis} - a system capable of reproducing and maintaining itself - mentioned above, and to the idea of  Autocatalytic sets \cite{Hordijk 2013} \cite{Hordijk 2018}. However I will not develop those links here. Rather my purpose is to claim that exactly the same applies in engineering systems in general, and even in physics in some cases. That is what I develop below. \\

Given that it is understood I am considering causal closure in terms of Levels and \textbf{LOIs}, I can summarise as follows:\footnote{This is related to the idea of inter level \textit{causal entanglement} \cite{Vecchi 2019}.}
\begin{quote}
	\textbf{Effective Causal Closure (\textbf{ECC}) in Biology}: \textbf{\textit{We have Effective Causal Closure of properties \textbf{P(d)} in a biological context when the considered set of levels \{\textbf{BL-TL}\} and data \{\textbf{HL-LL}\} is large enough to allow causal closure leading to autonomous biological functioning. It is ``Effective'' because (i) we know other levels do indeed have an influence, but can regard those influences as inputs to an autonomous system that do not destroy its autonomy, and (ii) it is a} ceteris parabus \textit{relation, as discussed in \S\ref{sec:open_controlled}. It can be destroyed by unpredictable Black Swan events \cite{Taleb 2010} that lie outside the normal operating environment.}}
\end{quote}   
Thus this characterizes the set of levels needed for an entity (a cell or an organism) to function successfully. There is then no preferred level enabling the system to function: they all equally enable this to happen \cite{Noble and Noble 2019 A-Mergence}. \cite{Green and Batterman} argue that in such cases, while the Effective Theories $\textbf{ET}_\textbf{L}$ are contained in the range \{\textbf{BL-TL}\}, one needs different models at each level:
\begin{quote}
	``\textit{No single mathematical model can account for behaviors at all spatial and temporal scales, and the modeler must therefore combine different mathematical models relying on different boundary conditions''  }
\end{quote} 
These are the different $\textbf{ET}_\textbf{L}$s for each level. But note that one then needs data $d_\textbf{L}$ for each level \textbf{L} too. Thus the set of levels where data is given has to be the same as the set of levels where \textbf{ECC} occurs. Thus \textit{\textbf{in the definition of Effective Causal Closure just given, one should set}} \{\textbf{BL-TL}\} = \{\textbf{LL-HL}\}. \cite{Green and Batterman} give the examples of epithelial sheets and mechanical modeling of gastrulation.\\

To be clear: one is free to work with an Effective Theory $\textbf{ET}_\textbf{L}$, with appropriate data for that level,  at any level chosen \textbf{L} in the range  \{\textbf{BL-TL}\}; but one only gets Effective Causal Closure by including that full set of levels and data.

\paragraph{When does this occur in biology?} There are two cases where \textbf{ECC} occurs in biology.
\begin{itemize}
	\item \textbf{Cells} The cellular level is the lowest level showing all the attributes of life. It is a case of \textbf{ECC} involving Levels \textbf{L4-L5}.
\item \textbf{Individuals} The organism level is the major coherent emergent level in life, assuming it is  a multicellular organism such as a human being. This is a case of \textbf{ECC} involving levels \textbf{L4-L7}.
\end{itemize}
However there is an interesting different view: that  human beings are essentially social beings, so that in fact it is a mistake to view them as being capable of living on their own, as is implied by that categorisation, Thus Berger and Luckmann \cite{Berger and Luckmann 1991} wrote about the \textit{Social Construction of Reality}: our worldview - an inescapable part of our nature shaping our actions - is crucially shaped by the society in which we live. Merlin Donald's book \textit{A Mind so Rare} \cite{Donald 2001} essentially agrees, as does Andy Clark's book \textit{Supersizing the Mind} \cite{Clark 2008}. In short, top-down effects from society so crucially shape our being that they are not just perturbations of independent existence: they are essential, and  that characterisation is wrong. The correct \textbf{ECC} statement is 
\begin{itemize}
	\item \textbf{Social human beings} Human beings are essentially social, and are in fact a  case of \textbf{ECC} involving levels \textbf{L4-L8}.
\end{itemize}
\paragraph{Ignoring the lower levels}A  further key comment regards the other end of the scale: why is it legitimate to ignore levels \textbf{L1-L3} here? The answer is the existence of \textit{quantum and classical protectorates} that are governed by emergent rules and are insensitive to microscopics \cite{Laughlin_Pines_2000}.  This is another way of affirming the causal efficacy of the Effective Theories $\textbf{ET}_\textbf{L}$ at each emergent level \textbf{L.} . However the  Effectively Causally Closed levels will reach down to determine what happens at those levels via time dependent constraints (\ref{eq:constraints}) \cite{Ellis_Kopel_2019}.

\begin{quote}
	\textbf{In summary:
	\textit{ Interlevel Effective  Causal Closure as identified here is a key feature of biological functioning,} \textit{emphasized in \cite{Bechtel 2007} and \cite{Moreno and Mosseo 2015}. As well as being key in terms of emergence \textbf{P(d)} of properties, it is also key in terms of  evolutionary and developmental processes \textbf{E(d)}, see \cite{Carroll 2005} (where it is not identified as such, but is there) and \cite{Ruiz-Mirazo and Moreno 2012}, where the relation is made explicit.}}
\end{quote} 

\paragraph{Setting the data}
One final issue remains: we don't in practice set data at all the levels \textbf{LL-HL} required in a particular context. 
\begin{quote}
	\textbf{Setting data}: \textbf{\textit{In practice one sets data at the highest level \textbf{HL=TL} that is relevant to a particular problem, and lets downward causation cascade data down to choose any set of data in the required equivalence class at each lower level down to the lowest level {LL=BL}}}
\end{quote}
There is no way we could in fact set the data at the lower levels. And there is no need for us to do so. This comment applies both in theory and in practice. That is, it is a statement both about epistemology (what we can know) and ontology (what we can do). 
\subsection{Inextricably Intertwined Levels:  Biology}\label{sec:Intertwined_Biology}
\paragraph{A higher level may be essential to a lower level}
An important possibility is that the properties \textbf{P(d)} of two levels \{\textbf{BL,TL}\} may  have an essential relationship with each other: each level cannot function without the other, as in some cases of \href{https://en.wikipedia.org/wiki/Symbiosis}{symbiosis}.  
\begin{quote}
	\textbf{Inextricably Intertwined Levels  \textit{Two levels \textbf{BL}, \textbf{TL} are inextricably intertwined levels  (\textbf{IIL}) if the effective dynamics $\textbf{ET}_{\textbf{BL}}$, $\textbf{ET}_{\textbf{TL}}$ at each of the two levels cannot occur without involving the other.}}
\end{quote}
Consider an individual  human being. It is no surprise that the level \textbf{L7} of the individual cannot exist without the level \textbf{L5} of cells, for human beings are made out of cells and depend on them for their existence and physiological functioning. 
But the fact is that the converse is also true: the cells cannot exist and function without the existence of the body that they comprise. The reason is that cells have specialised for specific functions, and cannot survive on their own. They are supplied with oxygen-laden blood by the lungs, heart, and indeed the entire \href{https://en.wikipedia.org/wiki/Circulatory_system}{circulatory system}, without which they die in a matter of minutes (as happens if a heart attack occurs). Thus levels \textbf{L5} and \textbf{L6} are inextricably intertwined. But  organs are part of the individual and won't function without systemic integration at that level. Hence levels \textbf{\{L5-L7\}} are in fact inextricably intertwined. \\

\noindent Now an interesting issue arises: \textbf{ECC} occurs for levels \textbf{L5-L7}. Should I have included \textbf{L4} in the inextricably intertwined levels? Certainly Level \textbf{L4} is required in order that cells exist at level \textbf{L5}, but is the other way round true also? I believe one can claim it is, because the  gene regulatory networks that control production of proteins at Level \textbf{L4} are at Level \textbf{L5}, and they would not exist if it were not for their functioning. Thus the real inextricably intertwined set of levels is \{\textbf{L4-L7}\}: the same as the \textbf{ECC} set of levels. There is however this difference: the \textbf{ECC} relation is \textit{ceteris parabus}, as explained above. The \textbf{IIL} relation is not, it is essential, whatever happens at other levels, these levels are crucially dependent on each other. 

\section{
	Interlevel Causal Closure: Digital Computers, Physics}\label{sec:causal_closure_computers}
The discussion in the last two sections makes clear a set of principles that apply equally to engineering, and that is what I will show in this section. 

To make
the discussion concrete, I will consider the case of digital computers.  But it will apply equally to other branches of engineering: automobiles, aircraft, chemical plant, water supply systems, sewerage systems, and so on.

I consider the nature of digital computers (Section \ref{sec:computers}), where Interlevel Effective Causal Closure again occurs (Section \ref{sec:Effective_computer}).  Inextricably Intertwined Levels again occur in computers (Section \ref{sec:Intertwined_computers}), and  in physics and chemistry (Section \ref{sec:Intertwined_phyiss}). 
\subsection{Digital computers}\label{sec:computers}
 The relevant hierarchy \cite{Tannenbaum} \cite{Ellis_Drossel_computers}\footnote{The labeling of levels is a bit different than in \textbf{Table  1} and \textbf{Table 3} because the focus here is specifically on computers.} is shown in \textbf{Table 4}.

\vspace{0.1in}

\begin{tabular}{|c|c|c|c|}
	\hline \hline
Level 	& Entity & Nature \\	
	\hline \hline
Level	\textbf{L9}& Global Society & Global Social and Economic Context      \\ \hline
Level	\textbf{L8} &  Country & Social and Economic Context      \\ \hline
	Level \textbf{L7}	& Internet &  Maximal Network \\ 
	\hline 
	Level \textbf{L6}	& Network &Linked computers, printers, file servers  \\ 
	\hline 
Level	\textbf{L5}	& Computer & Integrated Circuits,   I/O devices, Memory devices \\ 
	\hline 
Level	\textbf{L4} & Integrated circuits	& ALOE, CPU, Memory, linked by bus   \\ 
		\hline 
Level	\textbf{L3}	& Gates & Boolean logic: AND, OR, NOT   \\ 
	\hline 
Level	\textbf{L2}	& Transistors & Binary ON/OFF function    \\ 
	\hline 
Level	\textbf{L1}	& Crystalline structure & Symmetry, Band Structure    \\ 
	\hline 
Level	\textbf{L0}	& Electrons, Ions, Carriers & Structure, Current Flow   \\ 
	\hline \hline
\end{tabular} 

\vspace{0.1in}

\textbf{Table 4}:\label{Table2} \textit{Computer Implementation Hierarchy (schematic). This is the physical context within which upward emergence and downward causation takes place in the case of digital computers. For a full discussion, see Chapter 2 of \cite{Ellis_2016}. }

\paragraph{Upward and downward causation} 
The dynamics of a computer is driven by the algorithms encoded in the programs loaded, together with the data used by those programs. 
These control the flow of electrons through gates at the transistor level via a combinations of 
upward and downward causation (\cite{Ellis_2016}:Chapter 2), \cite{Ellis_Drossel_computers}. This enables the emergent effective laws $\textbf{ET}_\textbf{L}$ at each level \textbf{L}. Different algorithms result in different flows of electrons, as can be demonstrated by running different computer programs which produce different patterns of electron flows through transistors at Level \textbf{L2}, and cause major effects at social levels \textbf{L8} and \textbf{L9} \cite{MacCormick_2011_algorithms}.

\paragraph{The bigger picture} The fact that the higher levels \{\textbf{L8,L9}\} reach down to affect what happens at the lower levels \{\textbf{L0-L6}\} is stated in   \cite{Ellis_Drossel_computers} as follows:  
\begin{quote}
\textit{\textbf{Causal closure in the case of computers}: In the real world, it is only the combination of
physics with its logical, social, psychological, and engineering contexts
(which includes the values guiding policy) that can be causally complete,
because it is this whole that determines what computer programs will be
written and what data utilised, hence what electron flows will take place
in integrated circuits, as per the discussion in this paper}
\end{quote}
This is in parallel to the interlevel causal closure that takes place in biology, as discussed in Section \ref{sec:causal_closure_biology}. \cite{Ellis_Drossel_computers} gives a specific example:
\begin{quote}
	\textit{``As a specific example: the amount of money that can be dispersed to you from
an ATM will be limited by an agreement you have reached with your bank.
The program used to control the ATM will take into account the existence of
such limits, and the specific amount you are able to take out in a given time
period will be limited by a logical AND operation linking this agreed amount
to the amount of money in your account. Thus these abstract variables will
control electron flows in both the bank computers and the ATM dispenser
mechanism. Every relevant abstract variable has physical counterparts; in
other words, it?s realized by some physical properties on some relevant physical
substrate.''}
\end{quote} 
But crucially there is much more than this: there is the whole issue of the purposes computers are used for in society, from controlling manufacturing to enabling the internet, cell phones,and social media, and they way that this whole enterprise is shaped by the values of those that control the system. The book \textit{Coders} \cite{Thompson 2019} considers ``\textit{the morality and politics of code, including its implications for civic life and the economy. Programmers shape our everyday behavior: When they make something easy to do, we do more of it. When they make it hard or impossible, we do less of it}.'' All this is expressed in the flows of electrons through gates at the digital levels \textbf{L2-L3}.  
\subsection{Effective Causal Closure in Computers}\label{sec:Effective_computer}
Nevertheless, just  as in the case of biology, Effective Causal Closure can occur when interlevel causation results in a high degree of autonomy of operation. 
Analogously to the case of biology, one can state
\begin{quote}
	\textbf{Effective Causal Closure (\textbf{ECC}) in Computers}: \textbf{\textit{Effective Causal Closure  of properties \textbf{P(d)} in a digital computer occurs when the set of levels considered are large enough to allow causal closure leading to autonomous  functioning. It is ``Effective'' because (i) we know other levels do indeed have an influence, but believe we can regard those influences as inputs to an autonomous system that do not destroy its autonomy, and (ii) it is a} ceteris parabus \textit{relation, as discussed in \S\ref{sec:open_controlled}. It can be destroyed by Black Swan events that lie outside the normal operating environment.}}
\end{quote}   
\noindent The two emergent levels with their own causal integrity emerging through \textbf{ECC} are the \href{https://en.wikipedia.org/wiki/Integrated_circuit}{integrated circuit} level  (\textbf{L4}),  the equivalent of the cell in biology, with ECC given by Levels \textbf{L2-L4}; and the computer level (\textbf{L5}), the equivalent of the individual in biology, with ECC given by levels \textbf{L2-L5}. But just as in the case of biology one can make a case that one should really include the societal level, the same applies here too. The quotes above suggest that ECC for computers really only occurs for Levels \textbf{L2-L9}, including the highest level because of the effect of the World Wide Web. 

\paragraph{Engineering and Applied Physics} The same kind of considerations apply to all branches of engineering, and equally to all branches of applied physics. The applications (socially determined at Level \textbf{L8}) determine what physical effects occur (Levels \textbf{L1-L3}).

   \subsection{Inextricably Intertwined Levels:    Computers}\label{sec:Intertwined_computers}
   
 \paragraph{Inextricably Intertwined Levels (\textbf{IIL}s)} Do these occur in this case too, as they did in biology?
 Here there is a major difference: the transistors do not depend on the computer for their continued existence, whereas cells depend on the organism for their existence. While in biology \textbf{IILs} link the individual as a whole to the molecular level, here they also occur, but only at the levels \textbf{L0-L1} in \textbf{Table 4}.
 
 The reason for that relation is that downward emergence of key  properties at the electron level \textbf{L0} takes place, due to properties of the crystal level \textbf{L1}, as explained in detail in \cite{Ellis_2020_response}. This is called  a ``Foundational
 Determinative Relation'' (FDR) by Carl Gillett, see  \cite{Gillett}. In more detail, 
  \href{https://en.wikipedia.org/wiki/Quasiparticle}{quasiparticles} such as  \href{https://en.wikipedia.org/wiki/Phonon}{phonons} exist due to the broken symmetries of the emergent lattice structure. They come into being as effective particles at the lower level \textbf{L0} because they are dynamically equivalent to collective oscillations of a level \textbf{L1}  structure (the crystal lattice) (\cite{Simon 2013}:82-83), 
 
 This is an essentially quantum theory phenomenon.  One can think of it as an interlevel  \textbf{wave(macro)-particle(micro) duality.} \cite{Franklin_Knox_Phonons} say it this way:
 \begin{quote}
 	\textit{``Phonons [are] quasi-particles that have some claim to be emergent, not least because the way in which they relate to the underlying crystal is almost precisely analogous to the way in which quantum particles relate to the underlying quantum field theory.''}
 \end{quote}
 Stephen Blundell states the key point thus (\cite{Blundell 2019}:244):
 \begin{quote}
 	\textit{``So now we come to the key question: Are these emergent particles real? From the perspective
 		of quantum field theory, the answer is a resounding yes. Each of these particles emerges from a
 		wave-like description in a manner that is entirely analogous to that of photons. These emergent
 		particles behave like particles: you can scatter other particles off them. Electrons will scatter off
 		phonons, an interaction that is involved in superconductivity. Neutrons can be used to study the
 		dispersion relation of both phonons and magnons using inelastic scattering techniques. Yes, they
 		are the result of a collective excitation of an underlying substrate. But so are `ordinary' electrons  	and photons, which are excitations of quantum field modes.'}'
 \end{quote}
 As a consequence, the levels \{\textbf{L0,L1}\} are inextricably intertwined. Another way of stating this  is the way that \href{https://en.wikipedia.org/wiki/Bloch_wave}{Bloch's Theorem} \cite{Bloch 1929} shows how the crystal structure causes  the   electron wave functions to have a basis consisting of Bloch  eigenstates with the same periodicity as the crystal.

This is a proof  that the solid state physics occurring in digital computers is a case where causal closure is impossible at the micro level \textbf{L0} alone.  It  also shows that in general the set of \textbf{IILs} is not the same as the \textbf{ECCs}.

\subsection{Inextricably Intertwined Levels:   Physics and Chemistry}\label{sec:Intertwined_phyiss}
While these considerations apply to digital computers, of course they also apply in particular to the solid state physics itself that underlies their operation, due to the nature of crystals. For the reasons just discussed
\begin{quote}
	\textbf{In Solid State Physics, \textit{as a consequence of interlevel wave-particle duality, levels \{\textbf{L1,L4}\} in Table 1 are inextricably intertwined levels. }}
\end{quote}
However the \textbf{IIL} phenomenon is not confined to this case. The laser is another example (\cite{Pezzulo and Levin 2016}:\S2) 
\begin{quote}
	``\textit{The laser involves a kind of `circular' causality
which occurs in the continuous interplay between macrolevel
resonances in the cavity guiding, and being reinforced by,
self-organization of the molecular behaviour}''. 
\end{quote}
A quite different example is Resonance Energy Transfer (RET) in the transport of electronic energy from one atom or molecule to another \cite{Jones and Bradshaw 2019}. As described in that paper, 
\begin{quote}
	``\textit{The individual electrons do not migrate between molecules during the transfer process, since the molecular orbitals (the wavefunctions) do not overlap, but instead move between individual electronic states within the molecules. ... energy transfer, through dipole coupling between molecules, mostly depends on two important quantities: spectral overlap and intermolecular distance''}
\end{quote}
This lead to a $r^{-6}$ distance-dependence  for the resonance energy transfer rate in the short-distance regime. The behaviour results from interwining between the electron level \textbf{L1} and the molecular level \textbf{L4}. Because of these interactions,  Second-order perturbation theory is the minimal needed to describe RET. While this paper refers to `molecules', RET occurs in  atoms, chromophores, particles and carbon nanotubes. Its applications include nanosensors and photodynamic therapy.

This raises the issue that whenever molecular physics \cite{Buyanal 1997} is concerned, there is inextricable intertwining between the molecular structure at Level \textbf{L4}, bound by electrons, and the motions of the electrons at Level \textbf{L1}, controlled by that structure. This is manifest in binding energy (\cite{Buyanal 1997}:17-19) (\cite{McQuarrie_2008_quantum_chem}:569) and the existence of covalent bonds between atoms (\cite{Buyanal 1997}:17-19).
\begin{quote}
\textbf{Molecular physics 
	and quantum chemistry 
	\textit{also exhibit \textbf{IILs} between levels \textbf{L1 and \textbf{L4}}, which is why chemistry is a classic example of downward causation} \cite{Luisi}}.
\end{quote}
The group theory underlying this intertwining is discussed in \cite{Bishop_Ellis}.    
\section{Conclusion}\label{sec;conclude}
It is believed by many that because the bottom-most physics level is causally complete, and only upward causation takes place,  higher levels are purely derivative: they have no real causal validity. In this paper and its companion \cite{Ellis_2020_response}, I argue against that position. It is invalid because it treats physics in a way that ignores context, whereas physics outcomes always depend on context. 
In fact downward effects imply the opposite: in real world contexts, the bottom-most physics level is not by itself causally complete. 

In this section, I look at the contextual nature of causal closure of physics (Section 
\ref{sec:CC_context}), the way that unpredictability undermines causal closure of physics \textit{per se} (Section \ref{Sec:predictability}),
and comment on ways people ignore the issues discussed in this paper (Section \ref{sec:ignore}).  

\subsection{The Contextual Nature of Causal Closure of Physics}\label{sec:CC_context}
We can consider physics \textit{per se}, or  in relation to the natural world,  or in relation to biology, or  in relation to engineering.
Within physics, the issue of causal closure  depends on what aspects we are considering: Particle Physics, Nuclear Physics, Condensed Matter Physics, Cosmology for example.
Firstly, we have no reliable tested \textbf{TOE} at the very bottom level \textbf{L0}. We don't try to reduce  to that most fundamental physical level (\S\ref{sec:LOI_restricted}). Rather we reduce to a level that is convenient. That that can work is due to the existence of Quantum Protectorates, as explained in \cite{Laughlin_Pines_2000}.

 But then  it is common to assume that Level \textbf{L1} (particle physics) is causally complete. Is that indeed so? 
 I have argued that this is not the case in the contexts of 
physics and  biology (Section \ref{sec:causal_closure_biology}); 
physics and engineering, as exemplified by digital computers  (Section \ref{sec:Effective_computer});  solid state physics (Section \ref{sec:Intertwined_computers}); and physics and chemistry (Section \ref{sec:Intertwined_phyiss}). In each case the real causal closure that takes place is an interlevel affair, as emphasized in particular in the case of biology by many perceptive writers (Section \ref{sec:causal_closure_biology}). Effective Causal Closure in real world contexts spans many levels, in the case of biology reaching down from the level of the organism to the underlying physics via time dependent constraints.  
 This  implies how it works in terms of physics in relation to society. The causal effects of the corona virus pandemic at the social level reaches  down to cause major changes at the  physical levels \textbf{L1-L3} through a complex interaction between social behaviours, virology, and microbiology that for example has temporarily destroyed international air travel and so the trajectories of the billions of particles that make up aircraft. Causal closure only occurs when we take all these factors and levels into account.
 
Considering only disembodied physical laws seriously misleads about the nature of causation and causal closure in real world contexts.  
In summary, 
\begin{quote}
	\textbf{Causal closure of physics	\textit{In the real world context of engineering and biology, physics at the lowest level considered, whatever that is, is not  by itself causally complete. Interlevel causal closure involving engineering or biological variables, in those respective cases,  is required in order to have an effectively  causally closed system.}
}\end{quote}
This is formalized by the concepts of Effective Causal Closure (Sections  \ref{sec:causal_closure_biology} and  \ref{sec:Effective_computer}), and Inextricably Intertwined Levels (Sections  \ref{sec:Intertwined_Biology} and \ref{sec:Intertwined_computers}).
Within solid state physics itself, the lower levels \textbf{L1-L3} are not causally closed by themselves because of \textit{interlevel particle-wave duality} between the particle level \textbf{L1} where electrons and phonons live, and the crystal level \textbf{L4} where lattice vibrations take place  (Section \ref{sec:Intertwined_computers}). Properties of Level \textbf{L4} decouple from the lower physics levels.  As stated by \cite{Laughlin_Pines_2000},
\begin{quote}\textit{
	``The crystalline state is the simplest known example of a
	quantum protectorate, a stable state of matter whose generic
	low-energy properties are determined by a higher organizing
	principle and nothing else.''} 
\end{quote}

\subsection{Unpredictability undermines causal closure of physics \textit{per se}
}\label{Sec:predictability}
Ignoring interlevel issues, if by causal closure one means that data specified as precisely as possible leads to unique outcomes, then unavoidable unpredictability  undermines the possibility of physics \textit{per se} (whether quantum or classical) being causally closed. 

\paragraph{Quantum Physics} Quantum effects doubly cause uncertainty in outcomes. 

Firstly, the Heisenberg \href{https://en.wikipedia.org/wiki/Uncertainty_principle}{Uncertainty Principle} states that the standard deviations of position $\sigma_x$ and  momentum $\sigma_p$ obeys
\begin{equation}\label{eq:uncertainty}
\sigma_x \sigma_p \geq \hbar/2
\end{equation}
so one cannot even in principle apply Laplace's dream of setting initial data precisely at level \textbf{L1}. Consequently, outcomes are also uncertain.

Secondly, collapse of the wave function introduces an irreducible uncertainty in classical outcomes at this level when interactions take place \cite{Ghirardi}. 
This can reach up to macro levels through various amplifiers such as photon multipliers and CCDs. 
 In the engineering case it causes predictability issues at macro scales in terms of digital computer reliability because  \href{https://en.wikipedia.org/wiki/Cosmic_ray}{ cosmic rays}  cause errors in computer memories \cite{Cosmicray_compuer} \cite{Cosmicray_computer1}, and the emission of a cosmic ray by an excited atom is a quantum event that is unpredictable  even in principle. As regards biology, cosmic rays have had a significant effect on evolutionary history by causing genetic mutations \cite{Percival}. 

\paragraph{The Classical Case}
Uncertainty of outcomes occurs in this case too, because one can't set initial data to infinite precision \cite{Gisin_2019}. This is an outcome of the fact that infinity never occurs in physical reality \cite{Ellis Meissner Nicolai}. Thus physics is 
not causally closed in the classical case at higher levels because of chaotic dynamics (the butterfly effect), together with the impossibility of specifying initial data to infinite accuracy. This occurs for instance at the level at which fluid motion is determined.

 \cite{Anderson_01} 
puts it this way: 
\begin{quote}
	``\textit{A fluid dynamicist when studying the chaotic outcome of convection in a Benard cell knows to a gnat's eyelash the equations of motion of his fluid but also knows, through the operation of those equations of motion, that the details of the outcome are fundamentally unpredictable, although he hopes to get to understand the gross behaviour. This aspect is an example of a very general reality: the existence of universal law does not, in general, produce deterministic, cause-and-effect behaviour}''
\end{quote}
This fundamentally undermines the concept of a causally closed  physical levels in the case of classical physics. As was already known to Poincare, this occurs even in the \href{https://en.wikipedia.org/wiki/Three-body_problem}{3-body} gravitational case.

\paragraph{Microbiology} In the case of microbiology, interactions take place in the context of what Hoffmann \cite{Hoffman} has called ``The Molecular Storm''.  Molecular machines use ratchet-like mechanisms to harness energy from that storm, and organisms use it to provide an ensemble of options from which they can choose preferred lower level states and so attain biological objectives \cite{Noble and Noble 2018 Stochasticity} (see the quote in \S\ref{sec:upward_downward}). One has the opposite of the calm relation between initial data and outcomes supposed by Laplace. 

\subsection{How to ignore the issue}\label{sec:ignore}
Here are some ways that the nature of causal closure as discussed in this paper is avoided.

\paragraph{Partial reduction} This is very common. 

Francis Crick in  \href{https://en.wikipedia.org/wiki/The_Astonishing_Hypothesis}{\textit{The astonishing hypothesis}} \cite{Crick 1994} states,
\begin{quote} 
	\textit{``You, your joys and your sorrows, your memories and your ambitions, your sense of personal identity and free will, are in fact no more than the behavior of a vast assembly of nerve cells and their associated molecules. ''} 
\end{quote}
In other words, he is reducing \textbf{L7} to \{\textbf{L5-L4}\}. Now my physics colleagues who believe that all that matters is the particle interactions at level \textbf{L1} will just laugh and say, cells at Level \textbf{L4} and molecules at \textbf{L5} are nothing but particles interacting with each other. Thus Crick believes in the reality and effectiveness of causality at Levels \textbf{L4} and \textbf{{L5}} that
for example \cite{Hossenfelder_2019} and \cite{Greene 2020}, who believe that all causality resides at Level \textbf{L1}, clearly must deny. 

 So why did Crick emphasize causality at those levels? The answer of course is that  those were the levels at which he worked - and experienced the effectiveness of causality in terms of the interactions between entities (molecules, neurons) at those levels. 

From a strictly reductionist viewpoint, this is an illegitimate move. It is however fine if you accept Noble's Principle of Biological Relativity  \cite{Noble 2012}, as extended in \cite{Ellis_2020_response} and this paper: then causality is real at the levels \{\textbf{L4,L5}\} he studies. But that removes Crick's  justification for denying the reality of causation at Level \textbf{L7}.




\paragraph{Ignoring context}
Crucial contextual effects are simply ignored by some writers.  e.g.  \cite{Hossenfelder_2019}  \cite{Greene 2020}. 
 The view is ``\textit{You  are nothing but a bag of particles, it's just a matter of particles interacting via a known set of forces}''.  Context has nothing to do with it. The physicists holding this view all come from the particle physics/cosmology side, where this is to some extent true. Physicists from the solid state  physics side (the largest section of the physics community) do not hold this view, see e.g. \cite{Anderson_94} \cite{Simon 2013}. This leads to the large divide in the physics community between these two groups, as discussed by Sylvan Schweber \cite{Schweber 1993}.

\paragraph{Denying top-down causation } There is a frequent denial of the possibility of top-down causation, even though it occurs in physics and cosmology. In the latter case it occurs in the context of primordial nucleosynthesis in the early universe (\cite{Peter and Uzan 2013}:\S4.3) and structure formation in the later universe (\cite{Peter and Uzan 2013}:\S5), which are both dependent on the cosmological context at level \textbf{L10}. That is the reason that primordial element abundances on the one hand and matter power spectra and Cosmic Background Radiation angular power spectra on the other  can be used to place strong limits on the background model parameters \cite{Planck}, as discussed in (\cite{Ellis_2016}:275-277). If there were not such downward causation, this would not be possible. 

Top down causation clearly occurs in biology \cite{Campbell_1974} \cite{Fink_and_Noble}  \cite{Noble2008_Music}. A recent example is \cite{Pezzulo and Levin 2016}:  
\begin{quote}\textit{``Top-down approaches focus on system-wide states as causal
actors in models and on the computational (or optimality)
principles governing global system dynamics.''}
\end{quote}
That paper gives fascinating examples of downward causation from morphogenesis and regenerative medicine. 

Downward causation is obvious in subjects other than  particle physics and cosmology, and in particular in the functioning of the brain \cite{Kandel 1998} \cite{Kandel 2012} \cite{Ellis_2016} \cite{Ellis 2018}. For 
 example, our minds are  shaped by society, and shape society, including material outcomes in terms of architecture, engineering, art, and so on.  Society does not exist without mind, and mind does not exist without society \cite{Berger} \cite{Berger and Luckmann 1991} \cite{Donald 2001}.
That is the nature of interlevel causal closure.

Version: 2020/06/30.

\end{document}